\documentclass[preprint2,twocolumn]{aastex631}

\usepackage{amsmath,graphicx}	
\usepackage{bm}
\usepackage{multirow}
\usepackage{booktabs}
\shorttitle{Radio-interferometric imaging with R2D2}
\shortauthors{Aghabiglou et al.}
\graphicspath{{./}{fig/}}
\newcommand{\xb}{\ensuremath{\boldsymbol{x}}}
\newcommand{\yb}{\ensuremath{\boldsymbol{y}}}
\newcommand{\nb}{\ensuremath{\boldsymbol{n}}}
\newcommand{\rb}{\ensuremath{\boldsymbol{r}}}
\newcommand{\hb}{\ensuremath{\boldsymbol{{h}}}}
\newcommand{\bb}{\ensuremath{\boldsymbol{{b}}}}

\newcommand{\Nb}{\ensuremath{\boldsymbol{\mathsf{N}}}}

\newcommand{\Db}{\ensuremath{\boldsymbol{\mathsf{D}}}}
\newcommand{\Cb}{\ensuremath{\boldsymbol{\mathsf{C}}}}

\newcommand{\Phib}{\ensuremath{\boldsymbol{\Phi}}}
\newcommand{\thetab}{\ensuremath{\boldsymbol{\theta}}}
\newcommand{\betab}{\ensuremath{\boldsymbol{\beta}}}

\newcommand{\eC}{\mathbb{C}}

\newcommand{\eR}{\mathbb{R}}
\definecolor{forestgreen}{rgb}{0.13, 0.55, 0.13}
\definecolor{maroon}{rgb}{0.5, 0.0, 0.0}
\definecolor{cobalt}{rgb}{0.0, 0.28, 0.67}
\definecolor{coolblack}{rgb}{0.0, 0.18, 0.39}

\newcommand{\highq}[1]{\textcolor{forestgreen}{#1}} 
\newcommand{\lowq}[1]{\textcolor{maroon}{#1}} 
\begin{document}
\title{The R2D2 deep neural network series paradigm for fast precision imaging in radio astronomy}


\correspondingauthor{Yves Wiaux}
\email{y.wiaux@hw.ac.uk}

\author[0000-0001-6024-649X]{Amir Aghabiglou}
\affiliation{Institute of Sensors, Signals and Systems, Heriot-Watt University, Edinburgh EH14 4AS, United Kingdom}

\author[0009-0004-1056-5619]{Chung San Chu}
\affiliation{Institute of Sensors, Signals and Systems, Heriot-Watt University, Edinburgh EH14 4AS, United Kingdom}

\author[0000-0002-7903-3619]{Arwa Dabbech}
\affiliation{Institute of Sensors, Signals and Systems, Heriot-Watt University, Edinburgh EH14 4AS, United Kingdom}

\author[0000-0002-1658-0121]{Yves Wiaux}
\affiliation{Institute of Sensors, Signals and Systems, Heriot-Watt University, Edinburgh EH14 4AS, United Kingdom}

\begin{abstract}
Radio-interferometric (RI) imaging entails solving high-resolution high-dynamic range inverse problems from large data volumes. Recent image reconstruction techniques grounded in optimization theory have demonstrated remarkable capability for imaging precision, well beyond CLEAN's capability. 
These range from advanced proximal algorithms propelled by handcrafted regularization operators, such as the SARA family, to hybrid plug-and-play (PnP) algorithms propelled by learned regularization denoisers, such as AIRI. Optimization and PnP structures are however highly iterative, which hinders their ability to handle the extreme data sizes expected from future instruments. To address this scalability challenge, we introduce a novel deep-learning approach, dubbed ``\textbf{R}esidual-to-\textbf{R}esidual \textbf{D}NN series for high-\textbf{D}ynamic range imaging''. R2D2's reconstruction is formed as a series of residual images, iteratively estimated as outputs of Deep Neural Networks (DNNs) taking the previous iteration's image estimate and associated data residual as inputs. It thus takes a hybrid structure between a PnP algorithm and a learned version of the matching pursuit algorithm that underpins CLEAN. We present a comprehensive study of our approach, featuring its multiple incarnations distinguished by their DNN architectures. We provide a detailed description of its training process, targeting a telescope-specific approach. R2D2's capability to deliver high precision is demonstrated in simulation, across a variety of image and observation settings using the Very Large Array (VLA). Its reconstruction speed is also demonstrated: with only few iterations required to clean data residuals at dynamic ranges up to $10^5$, R2D2 opens the door to fast precision imaging. R2D2 codes 
are available in the \href{https://basp-group.github.io/BASPLib/}{BASPLib}
library on GitHub.
\end{abstract}



\section{Introduction}
Astronomical imaging is a cornerstone of our quest to deepen our understanding of the Universe. It allows us to decipher the origins of galaxies, explore planetary systems, and probe the fundamental laws governing the cosmos. Modern radio telescopes, such as the MeerKAT \citep{jonas2016proc}, the Low-Frequency Array \citep[LOFAR;][]{van2013lofar}, and the Australian Square Kilometre Array Pathfinder \citep[ASKAP;][]{hotan2021},
have transformed the field, offering unprecedented resolution and sensitivity. However, these powerful instruments bring scalability and precision challenges. The data volumes they provide are vast and the expected dynamic range of the radio images to be formed spans multiple orders of magnitude. 

Aperture synthesis by radio interferometry captures incomplete Fourier information of the radio sky. Addressing the underlying ill-posed inverse problem calls for advanced image formation algorithms to deliver the science objectives. Five decades since its inception, the CLEAN algorithm~\citep{hogbom1974aperture} remains the standard in RI imaging, owing to its simplicity and computational efficiency. However, CLEAN reconstructions exhibit intrinsic limitations, particularly as their angular resolution is restricted to the nominal instrumental resolution. Complex emission with extreme dynamic ranges can also be challenging to CLEAN in spite of its advanced variants \citep[e.g.,][]{bhatnagar2004scale,cornwell2008multiscale,offringa2014}. Furthermore, manual intervention is often required for stability, due to the greedy nature of the algorithm. 

Computational imaging techniques from Bayesian inference \citep{junklewitz2016resolve,cai2018uncertainty,arras2019unified} and optimization theories \citep[e.g.,][]{wiaux2009compressed,carrillo2012sparsity,garsden2015,dabbech2015moresane,repetti2020forward} were developed over the last decade or so. Bayesian methods provide quantification of the uncertainty about the image estimate. Optimization algorithms enable injecting handcrafted regularization into the data. One such example is the SARA family~\citep{carrillo2012sparsity,onose2016scalable,onose2017accelerated,repetti2020forward,terris2022,terris2023plug}, promoting nonnegativity and average sparsity in a redundant wavelet dictionary. The SARA algorithm and its extension to unconstrained minimization, uSARA, demonstrated high imaging precision on real RI data from modern telescopes, outperforming CLEAN \citep{dabbech2018,dabbech2022,wilber23a}. However, all these methods are highly iterative, which hinders their scalability to the sheer data volumes expected from next-generation instruments such as the flagship Square {Kilometre Array \citep[SKA; ][]{Labate2022,Swart22}}. 

Recent advances in deep learning, from PnP algorithms to advanced end-to-end DNNs, have opened the door to a whole new paradigm in computational imaging owing to their modeling power and speed. PnP algorithms are at the intersection of optimization theory and deep learning. They entail training a denoising DNN, which is then deployed within an optimization algorithm as a substitute for a handcrafted, and often sub-iterative regularization operator \citep{venkatakrishnan2013,Chan2017,Zhang2017,Kamilov23}. One such example is AIRI, standing for ``AI for Regularization in radio-interferometric Imaging'' \citep{terris2022}, underpinned by the Forward-Backward algorithmic structure from convex optimization theory. AIRI has been demonstrated to achieve slightly superior imaging precision to uSARA at a lower computational cost on real large-scale high-dynamic range RI data from SKA pathfinder and precursor instruments \citep{dabbech2022,wilber23b}. 
Yet, PnP algorithms remain as highly iterative as their pure optimization counterparts. Fully data-driven end-to-end DNNs have been proposed \citep{connor2022deep}. They are able to provide ultra-fast reconstruction, but simultaneously lose robustness (interpretability and generalizability) by failing to enforce fidelity to data. More advanced end-to-end unrolled DNN architectures have emerged \citep[see][and references within]{Monga21}. Unrolled DNNs are model-based architectures designed to ensure the consistency of the reconstructed images with the measurements by unrolling the iteration structure of an optimization algorithm in its layers. They address the robustness issues of data-driven end-to-end approaches, and have been demonstrated to deliver a high degree of precision across a variety of applications \citep{Kamilov23}. However, they reintroduce a lack of scalability, owing to the fact that embedding large-scale measurement operators in the network architecture is impractical both during training and inference. This constraint is particularly acute for RI imaging.

To address the scalability challenge of PnP approaches, while preserving their robustness, we introduce a novel deep-learning approach, dubbed ``\textbf{R}esidual-to-\textbf{R}esidual \textbf{D}NN series for high-\textbf{D}ynamic range imaging''. R2D2's reconstruction is formed as a series of residual images, iteratively estimated as outputs of DNNs taking the previous iteration's image estimate and associated data residual as inputs. It thus takes a hybrid structure between a PnP algorithm and a learned version of the matching pursuit algorithm that underpins CLEAN.
The main contribution of this paper is twofold. Firstly, we provide a detailed description of the R2D2 algorithm, generalizing from recent work \citep{aghabiglou2023a,Hauptmann2018}. We discuss two incarnations distinguished by their DNN architectures. We also present a telescope-specific training methodology for R2D2. Secondly, we present an in-depth analysis of the algorithm's imaging precision and computational efficiency in simulation in comparison with the optimization algorithm uSARA, the PnP algorithm AIRI \citep{terris2022}, and CLEAN \citep{offringa2014,offringa2017}. In a sister article, we demonstrate the R2D2 algorithm trained for VLA on real high-dynamic range RI data \citep{dabbech2024}.

The remainder of this paper is as follows. Section \ref{sec:R2D2-algorithm} describes the R2D2 algorithm featuring its two incarnations. Section \ref{sec:Training-approach} elaborates on the training methodology of the R2D2 algorithm, including the underpinning core DNN architecture, the diversified ground truth database, and resulting VLA-focused training datasets. It also presents implementation details of the training and its computational aspects. 
Section \ref{sec:simulations-results} provides a detailed analysis of the R2D2 algorithm performance, in comparison with state-of-the-art RI algorithms. Conclusions and future work are stated in Section \ref{sec:Conclusions}. 

\section{R2D2 algorithm} \label{sec:R2D2-algorithm}
In this section, we recall the RI data model including its Fourier and image-domain formulations. We present the R2D2 iteration structure and explain the sequential training of its underpinning DNN series. We finally describe two different incarnations of the R2D2 algorithm relying on distinct DNN architectures. 

\subsection{Data model} \label{subsec:data-model}
In aperture synthesis by radio interferometry, at a given time instant, each pair of antennas acquires a noisy complex measurement, termed visibility, corresponding to a Fourier component of the sought radio emission \citep{thompson2017interferometry}. The collection of the sensed Fourier modes accumulated over the total observation period forms the Fourier sampling pattern, describing an incomplete coverage of the {2-dimensional} (2D) Fourier plane. With no loss of generality, let ${\xb^{\star}}\in\mathbb{R}_{+}^N$ represent the unknown radio image of interest. The RI data ${\yb}\in\eC^M$ can be modeled as:
\begin{equation}
{\yb}={\Phib} {\xb^{\star}}+\nb,
\label{eq:observation}
\end{equation}
where $\nb \in \eC^{M}$ represents a complex Gaussian random noise with a standard deviation $\tau>0$ and mean 0. The linear operator ${\Phib}\colon \eR^N \to \eC^M$ is the measurement operator (i.e., describing the acquisition process), consisting in a nonuniform Fourier sampling. The operator ${\Phib}$ can also include a data-weighting scheme to improve the observation's effective resolution \citep[e.g., Briggs weighting; ][]{briggs1995}. Performing a discrete Fourier transform with the expected large amount of data is impractical. Often, the incomplete Fourier sampling is modeled using the nonuniform fast Fourier transform \citep[NUFFT;][]{Fessler2003,thompson2017interferometry}. In our proposed algorithm, we adopt the image-domain formulation of the RI data, obtained from \eqref{eq:observation} via a normalized back-projection such that:
\begin{equation}
\xb_{\textrm{d}} =\kappa \text{Re}\{\Phib^\dagger \yb\}=\Db{{\xb^{\star}}}+ {\bb},
\label{eq:backprojection}
\end{equation}
where $\xb_{\textrm{d}} \in \eR^N$ stands for the back-projected data, known as the \textit{dirty} image, with $(.^\dagger)$ denoting the adjoint of its argument operator, and $\text{Re}\{\cdot\}$ the real part of its argument. Considering $\bm{\delta} \in \eR^N$, the image with value 1 at its center and 0 elsewhere, the normalization factor $\kappa>0$ is equal to $1/\max(\textnormal{Re}\{{\Phib}^{\dagger}{\Phib}\bm{\mathsf{\delta}}\})$, ensuring that the point spread function (PSF), defined as $\hb=\kappa \text{Re}\{{\Phib}^{\dagger}{\Phib}\bm{{\delta}}\} \in \eR^{N}$, has a peak value of 1. The linear operator $\Db \colon \eR^N \to \eR^N $ maps the unknown image of interest to the dirty image space by encoding the Fourier de-gridding and gridding operations, and is defined as $\Db \triangleq \kappa \text{Re}\{{\Phib}^{\dagger}{{\Phib}}\}$. Finally, the image-domain noise vector ${\bb} =\kappa \text{Re}\{{\Phib^\dagger \nb}\} \in \eR^{N}$ represents the normalized back-projected noise.

\subsection{Algorithmic structure} \label{subsec:Algorithmic-structure}
\begin{figure*}
\centering
\includegraphics[width=0.85\textwidth]{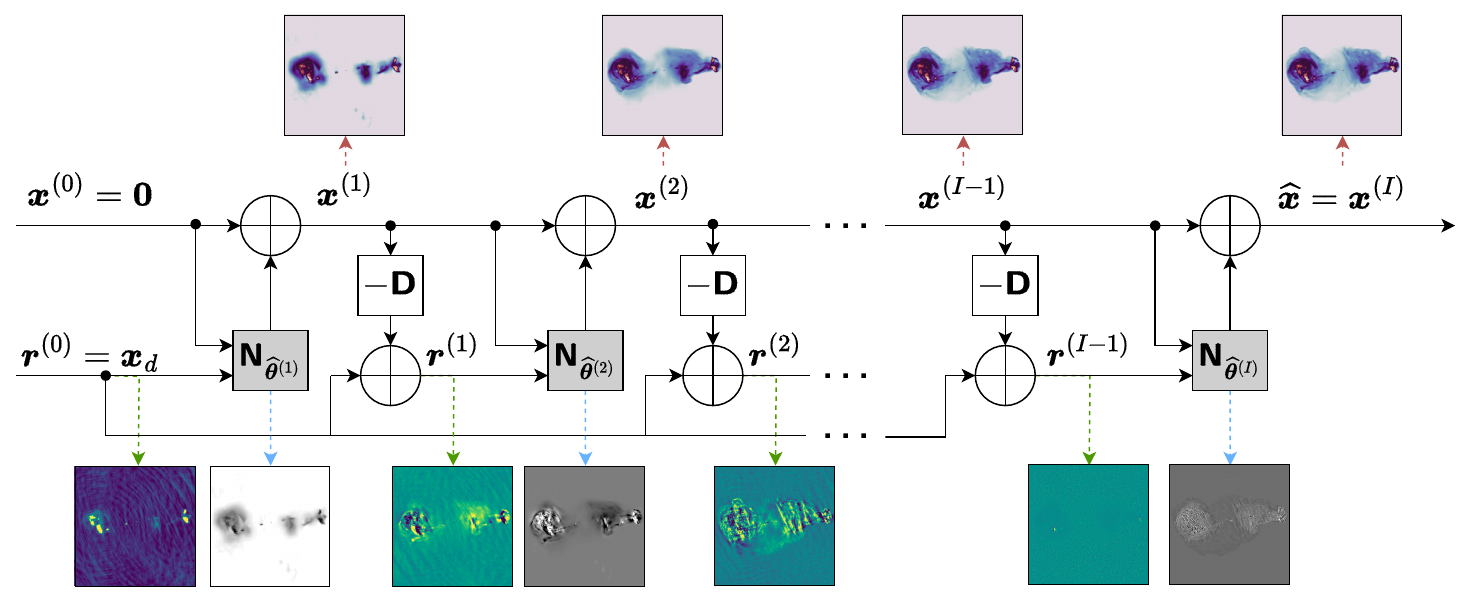}
\caption{
Illustration of the R2D2 algorithm. Both image iterates $\xb^{(i-1)}$ and associated residual dirty images $\rb^{(i-1)}$ are fed to R2D2 DNNs as input. R2D2 DNNs' output are then used to update the next image iterates. The sequence of the image iterates and corresponding residual dirty images are indicated with dashed red and green arrows, respectively. The sequence of the learned residual images are indicated with blue arrows. 
}
\label{fig:r2d2_illust}
\end{figure*}

The R2D2 algorithm requires training a series of $I$ DNNs represented as $(\Nb_{\widehat{\thetab}^{(i)}})_{1 \leq i \leq I}$, and characterized by their learned parameters $({\widehat{\thetab}^{(i)}} \in \eR^{Q})_{1 \leq i \leq I}$. Each network component takes two input images consisting of the previous image estimate $\xb^{(i-1)}$ and its associated back-projected data residual $\rb^{(i-1)}$, to which we refer as residual dirty image given by:
\begin{equation}
 \rb^{(i-1)}=\xb_{\textrm{d}}-\Db\xb^{(i-1)}.
 \label{eq:residual_update}
\end{equation}
The current image estimate is updated from the output of the network component ${\Nb_{\widehat{\thetab}^{(i)}}}(\rb^{(i-1)}, \xb^{(i-1)})$. The iteration structure of the R2D2 algorithm reads: 
\begin{equation}
\xb^{(i)}=\xb^{(i-1)} + {\Nb_{\widehat{\thetab}^{(i)}}}(\rb^{(i-1)}, \xb^{(i-1)}),
\label{eq:image_update}
\end{equation}
with the image estimate and the residual dirty image initialized to $\xb^{(0)} = \bm{0}$ and $\rb^{(0)} = \xb_{\textrm{d}}$. 
R2D2's iteration structure enables progressive improvement of both the resolution and dynamic range of the image estimate. Illustration of the R2D2 algorithm is provided in Figure~\ref{fig:r2d2_illust}. One can see how the learned residual images depict smooth structure at early iterations, and progressively capture finer details together with fainter structure at later iterations. R2D2's reconstruction $\widehat{\xb}$ thus takes the simple series expression below:
\begin{equation}
\widehat{\xb} \triangleq \xb^{(I)}= \sum_{i=1}^{I} {\Nb_{\widehat{\thetab}^{(i)}}(\rb^{(i-1)},\xb^{(i-1)})}.
\label{eq:sum_interp}
\end{equation}
\noindent
Formally, $I$ is an algorithmic parameter with an appropriate procedure necessary to determine its value.

The R2D2 algorithm features a hybrid structure between a learned version of matching pursuit, of which CLEAN is a well-known example, and a Forward-Backward PnP algorithm, such as AIRI. All three algorithmic structures are iterative, alternating at each iteration between the computation of a residual dirty image and a regularization step. They differ by the way their regularization operators are built and the variables they take as inputs. Firstly, as per \eqref{eq:image_update}, the R2D2 DNNs take the current 
image estimate and its associated residual dirty image as two separate input channels, and are trained to output a residual image serving as an additive update to the current image estimate. Secondly, at each iteration, CLEAN identifies model components by projecting the residual dirty image onto a sparsity dictionary, either the identity basis in the standard CLEAN \citep{hogbom1974aperture,schwab1984}, or bespoke multiscale kernels in multiscale CLEAN~\citep{cornwell2008multiscale}. Its iterations follow the exact same structure as \eqref{eq:image_update}, with a minor cycle in lieu of R2D2's DNN. In contrast to an R2D2 DNN, CLEAN's minor cycle structure is iterative itself, taking information from the residual dirty image only (not explicitly the current image estimate). Thirdly, AIRI's image update is computed as the output of a learned denoiser, taking as input a linear combination of the current image estimate and its associated residual dirty image \citep{terris2022}.

\subsection{DNN series training}\label{subsec:DNN-series-training}
The R2D2 algorithm is underpinned by a series of DNNs, trained sequentially. The first network in the series {$\Nb_{\widehat{\thetab}^{(1)}}$} is trained using a dataset composed of $L$
ground truth and dirty image pairs $(\xb_l^\star,{\xb_{\textrm{d}}}_l)_{1\leq l\leq L}$, with the input image estimates initialized such that ${\xb_l^{(0)}}=\bm{0}$, for all $ 1\leq l \leq L$. Subsequent networks $\Nb_{\widehat{\thetab}^{(i)}}$, at any given iteration $i >1$, are trained using a dataset consisting of the image triplets $(\xb_l^\star,{\xb_l^{(i-1)}},{\rb_l^{(i-1)}})_{1\leq l\leq L}$. 
The learnable parameters ${{\widehat{\thetab}}{^{(i)}}}$ of each network are estimated by solving the loss function of the form:
\begin{equation}
\label{eq:r2d2_loss}
 \min_{{\thetab}^{(i)}\in \eR^Q}\frac{1}{L} \sum_{l=1}^{L} ~ \| {\xb}^{\star}_{l} - [{\xb}^{(i-1)}_{l} + {{\Nb}}_{{\thetab}^{(i)}}(\rb_{l}^{(i-1)}, {\xb}^{(i-1)}_{l})]_{+} \|_{1},
\end{equation}
where $\lVert.\rVert_{1}$ stands for the $\ell_1$-norm, and $[.]{+}$ denotes the projection onto the positive orthant $\eR^N_+$ to enforce nonnegativity of the image estimates throughout the iterative process. The output of each trained DNN is used to update the image estimates $({\xb}^{(i)}_{l})_{1\leq l\leq L}$ as specified in \eqref{eq:image_update}, and the associated residual dirty images $({\rb}^{(i)}_{l})_{1\leq l\leq L}$ as per \eqref{eq:residual_update}. The updated image pairs are then incorporated in the training dataset of the next DNN. The loss functions of the form \eqref{eq:r2d2_loss} are solved using the Root Mean Square Propagation (RMSProp) algorithm, with the learnable parameters of each network initialized from the estimated parameters of the preceding trained network.

The sequential training concludes when the evaluation metrics of the reconstruction quality achieved by the validation dataset reach a point of stabilization. This offers a way to determine the number of required networks $I$ at the training level, making R2D2 parameter-free when it comes to reconstructing an image from specific data.

\subsection{Normalization procedures}
To avoid generalizability issues arising from large variations in pixel value ranges between the training dataset and images of interest (e.g., test images), normalization procedures must be deployed. We consider an iteration-specific normalization at both training and inference stages. On the one hand, the training of the network component at any iteration $i>1$ takes as input a normalized dataset where each image triplet $(\xb_l^\star,{\xb_l^{(i-1)}},{\rb_l^{(i-1)}})$ is divided by $\alpha^{(i-1)}_l$, the mean value of the previous image estimate ${\xb_l^{(i-1)}}$. The first network component ($i=1$) is also trained using a normalized dataset, where each image pair $(\xb_l^\star,{\xb_{\textrm{d}}}_l)$ is divided by $\alpha^{(0)}_l$, the mean value of the dirty image ${\xb_{\textrm{d}}}_l$. On the other hand, at inference (see \eqref{eq:image_update} and \eqref{eq:sum_interp}), all network components are applied on normalized inputs, with their outputs de-normalized accordingly. Hence, formally: $ {\Nb} \mapsto \alpha {\Nb}(./\alpha)$, where $\alpha \in \eR$ is an iteration-specific normalization factor obtained via the exact same procedure as in the training stage.

\subsection{R2D2 incarnations} \label{subsec:R2D2-incarnations}
We present two incarnations of the R2D2 algorithm distinguished by the architecture of their network components $(\Nb_{\widehat{\thetab}^{(i)}})_{1 \leq i \leq I}$. The first features a fully data-driven end-to-end DNN architecture. The second features a data model-informed DNN architecture unrolling the R2D2 algorithm itself, dubbed R2D2-Net. 
Nonetheless, both incarnations rely on the same DNN core architecture denoted by $\Cb_{\betab}$, with ${\betab}\in \eR^{P}$ standing for its learnable parameters.

The first incarnation takes simply the DNN $\Cb_{\betab}$ as the architecture of its network components. Under this consideration, each network component takes the form $\Nb_{\widehat{\thetab}^{(i)}} \equiv \Cb_{\betab^{(i)}}$, and its learned parameters are given by $\widehat{\thetab}^{(i)}= \betab^{(i)}\in \eR^{Q=P}$, for any $1\leq i \leq I$. For simplicity, we refer to this incarnation as R2D2.

For the second incarnation, each of its R2D2-Net components comprises $J$ layers of the core architecture $\Cb_{\betab}$ interleaved with $J-1$ approximate data-fidelity layers. The learned parameters of each R2D2-Net component $\Nb_{\widehat{\thetab}^{(i)}}$ thus take the form $\widehat{\thetab}^{(i)} = [\betab_1^{(i)},\dots, \betab_J^{(i)}] \in \eR^{Q=P \times J}$. Specific to the data-fidelity layers, the residual dirty images are computed using the image-domain convolution with the PSF instead of the RI mapping operator $\Db$. This mapping is fast and memory efficient, enabling the training of R2D2-Net on GPUs. In reference to its nesting structure, we dub this incarnation of the R2D2 algorithm as the ``\textbf{R}ussian \textbf{D}oll \textbf{R2D2}'', in short R3D3.
\section{Training approach}\label{sec:Training-approach}
This section introduces the training methodology for both incarnations of the R2D2 algorithm. It covers the underpinning core architecture and explains the procedure to build training datasets from a database of low-dynamic range images, using VLA sampling patterns. Implementation details including the computational cost of the training are also provided.

\subsection{U-Net core architecture}\label{subsec:UNet-core-architecture}
We consider the well-known U-Net \citep{Ronneberger15} as $\Cb_{\betab}$, the core architecture of both incarnations of the R2D2 algorithm. R2D2 network components simply take the U-Net architecture. For R3D3, its R2D2-Net network components take the same U-Net in their $J$ network layers.

{Typically, a U-Net consists of two distinctive parts. The first is the contracting path which aims to capture the semantic aspects of the network's input, encapsulated in the so-called feature maps \citep{lecun2015deep}. To do so, it applies successive multi-channel convolutions with increasing number of channels, followed by pooling layers to reduce the spatial dimension. The second is the expanding path which mirrors the structure of the contracting path. It aims to restore the spatial resolution by applying successive multi-channel convolutions and up-sampling layers to the feature maps to increase their spatial dimension while integrating information from the contracting path.}

\begin{figure*}
\includegraphics[width=0.99\textwidth,trim={0 2.5cm 0 0},clip]{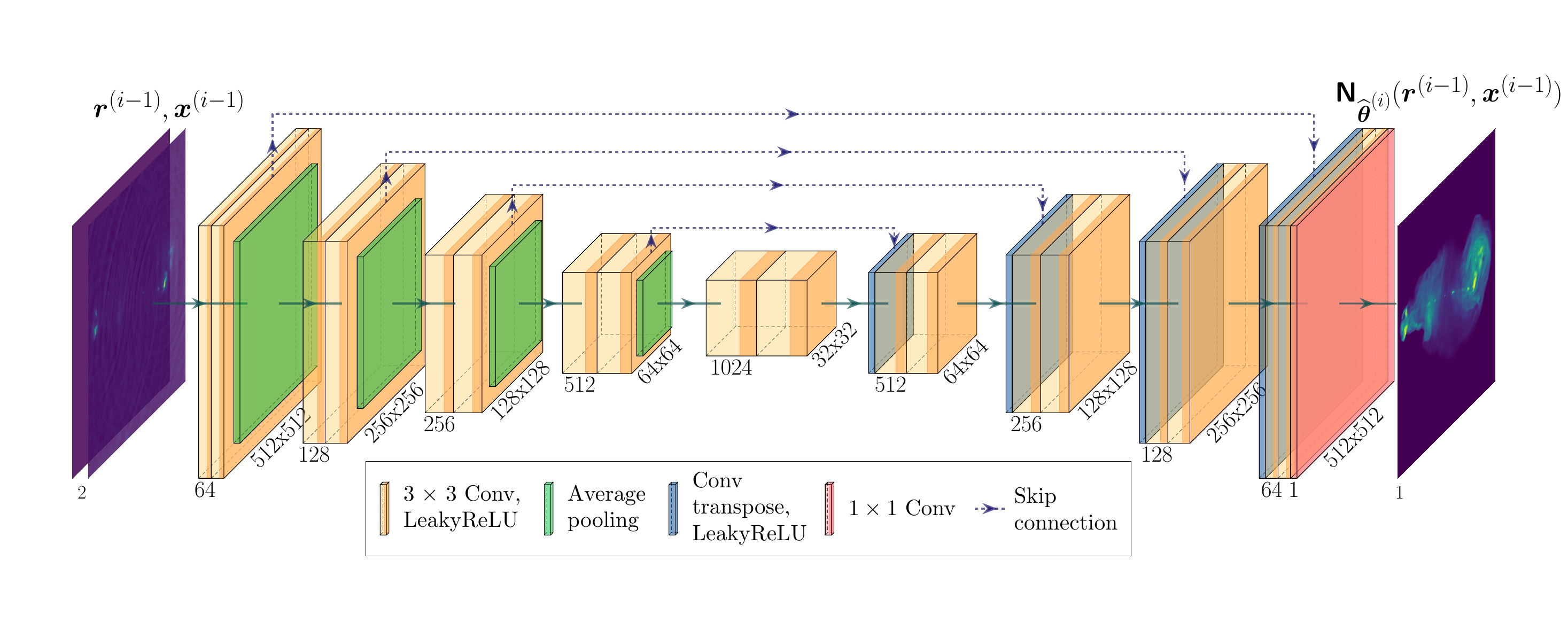}
 \caption{Illustration of the U-Net core architecture underpinning the different incarnations of the R2D2 algorithm. {The convolutional layers of the network, represented as boxes, apply multi-channel convolutions, followed by down-sampling (contracting path) or up-sampling (expanding path). The outcome of these layers are 3D feature maps, whose dimensions are specified around the corresponding boxes: the 2D spatial dimension at the bottom center, and the number of feature channels at the outer edge}.}
 \label{fig:unet_architecture}
\end{figure*}

{The adopted U-Net architecture is illustrated in Figure~\ref{fig:unet_architecture}. In the contracting path, each convolutional layer consists of two blocks of multi-channel $3\times3$ convolutions combined with the LeakyReLU activation function. Every two convolutional layers are followed by a 2D average-pooling layer with a stride of 2. After each down-sampling step, the number of feature channels is doubled. The first convolutional layer takes the network's input corresponding to the previous image estimate and its associated back-projected data residual, both of size 512$\times$512, and produces a 3D feature map of size $512 \times 512 \times 64$, with 64 being the number of feature channels. At the end of the contracting path, the reached bottleneck layer produces low-resolution 3D feature maps of size $32 \times 32 \times 1024$. The expanding path substitutes the 2D average-pooling with the 2D transposed convolution to double the spatial dimension of the output feature maps and utilizes skip connections from the down-sampling layers to restore the fine details. The final output layer employs a 1$\times$1 convolution layer resulting in a single-channel output that is the learned residual image.}

\subsection{Ground truth database}\label{subsec:Ground-truth-database}
{Key to the training of our deep-learning model is a large database of ground-truth images with a wide variety of features and dynamic ranges, and free from noise and artifact structure. In the absence of a large physical RI database readily providing these characteristics, we build a database of curated ground-truth images from real low-dynamic range astronomical and medical images, sourced as follows.
Radio astronomy images} are gathered from the National Radio Astronomy Observatory (NRAO) Archives, and LOFAR surveys, namely, LOFAR HBA Virgo cluster survey \citep{edler2023} and LoTSS-DR2 survey \citep{shimwell2022}. Optical astronomy images are gathered from the National Optical-Infrared Astronomy Research Laboratory. Medical images are selected from the NYU fastMRI Initiative Database \citep{zbontar2018fastmri,knoll2020fastmri}.
{Training using a curated database originating from other modalities and applications has shown to be effective for RI imaging \citep{terris2022,terris2023plug}}.

\begin{figure}
 \centering
 \setlength\tabcolsep{1pt} 
 \begin{tabular}{cccc}
 \includegraphics[width=0.24\linewidth]{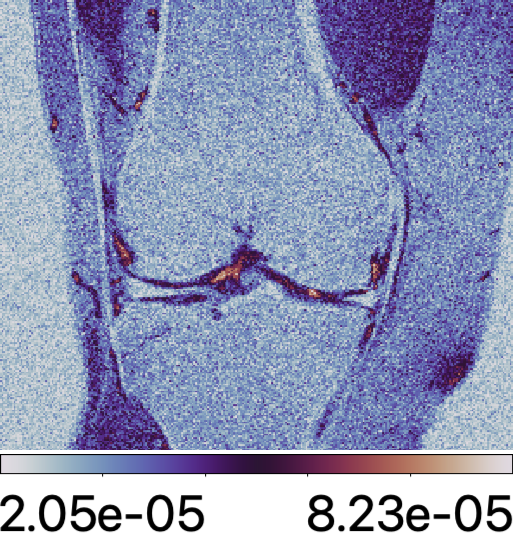} &
 \includegraphics[width=0.24\linewidth]{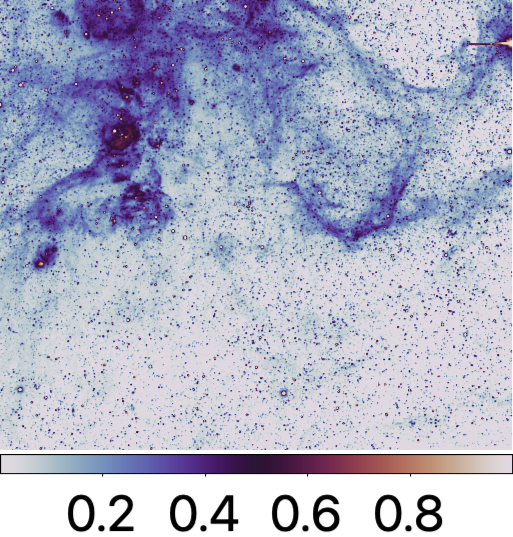} &
 \includegraphics[width=0.24\linewidth]{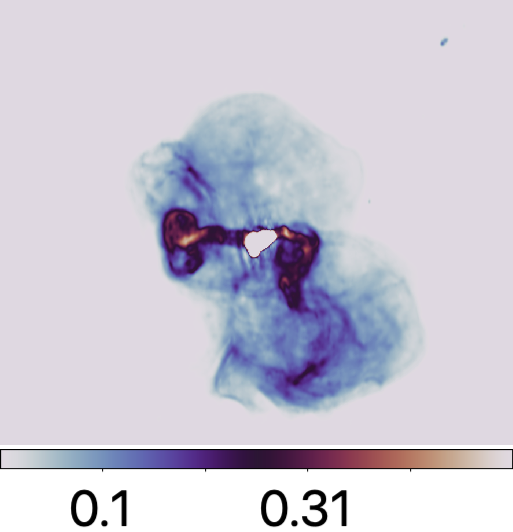} &
 \includegraphics[width=0.24\linewidth]{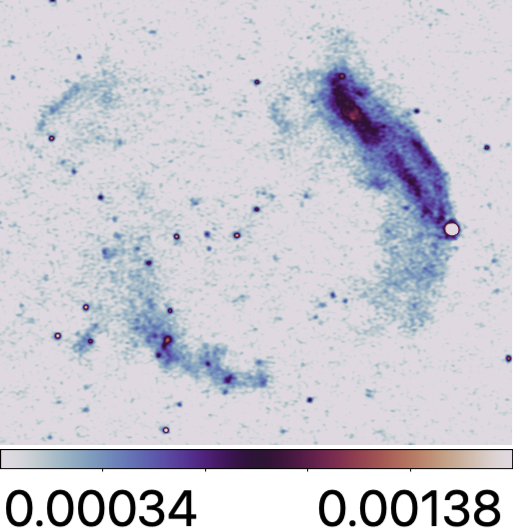}\\
 \includegraphics[width=0.24\linewidth]{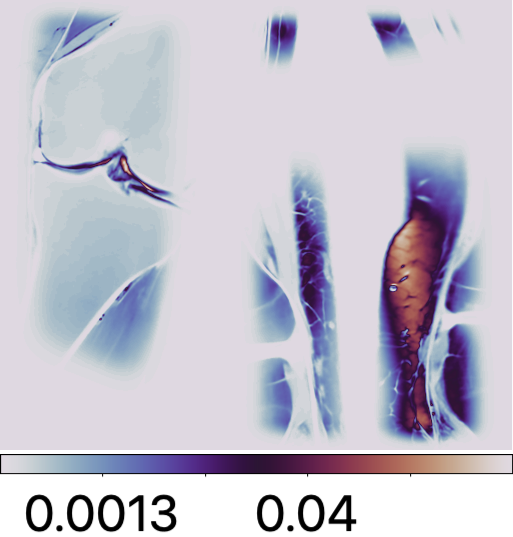} &
 \includegraphics[width=0.24\linewidth]{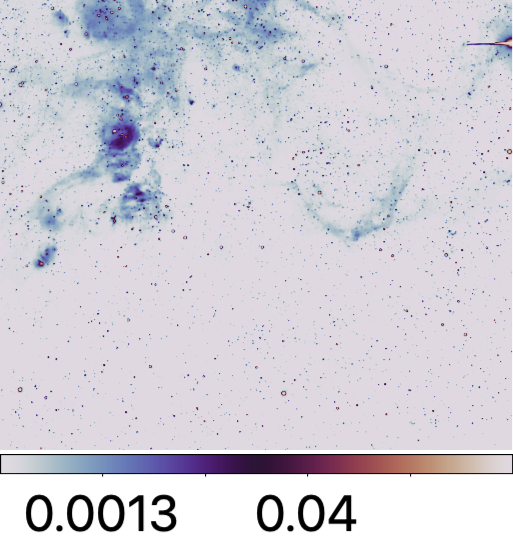} &
 \includegraphics[width=0.24\linewidth]{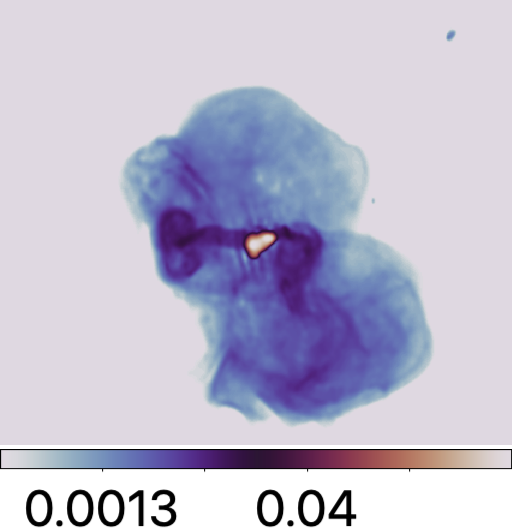} &
 \includegraphics[width=0.24\linewidth]{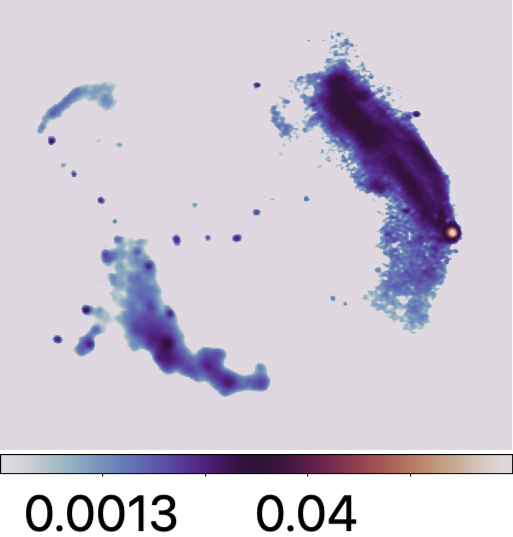}
 \end{tabular}
 \caption{Selection of raw low-dynamic range images (linear scale; top row), and corresponding ground-truth images after pre-processing (logarithmic scale; bottom row). The training dataset includes {medical images (first column), and optical astronomy images (second column)}. The validation dataset comprises images of giant radio galaxies (e.g., Messier 87; third column), and radio galaxy clusters (e.g., Abell~746; fourth column).\vspace*{-0.2cm}} 
 \label{fig:train_val}
\end{figure}

{Ground-truth images of size $N=512\times 512$,} are generated using the pre-processing procedure proposed in \citet{terris2023plug}. More precisely, various operations including concatenation, rotation, translation, and edge smoothing, are applied specifically to the medical images to deconstruct their anatomical features. Denoising is applied to all images, of both medical and astronomical origins, to eliminate artifacts and noise, using a denoising DNN \citep{zhang2023practical} in combination with soft-thresholding operations. Additionally, a pixel-wise exponentiation transform can be applied to the curated ground-truth images to emulate the characteristic high dynamic range of radio images \citep{terris2022}. Examples of raw low-dynamic range images and their corresponding denoised and exponentiated ground-truth images are shown in Figure~\ref{fig:train_val}. 
 
\subsection{VLA-specific training} \label{subsec:VLA-specific-training}
In training both incarnations of the R2D2 algorithm, we take a telescope-specific approach, encompassing all observation settings of a single telescope. While the methodology is implemented for the VLA, a similar process can be undertaken for other telescopes as well. 

VLA antennas are re-configurable into four configurations A, B, C, and D. For a given observation frequency, configuration A provides the longest baselines, hence the highest angular resolution. In contrast, configuration D provides the shortest baselines, {offering the best surface brightness sensitivity to extended emission}. Depending on the science target, astronomers often probe the radio sky with multiple configurations of the VLA. Herein, we opt for the combination of configurations A and C for a balanced Fourier coverage. To obtain diversified datasets, we consider a wide variety of observation setting, by uniformly randomizing (i) the pointing direction with a declination in the range [5,~60]~degrees and a right ascension in the range [0,~23]~hr~(J2000), (ii) the total observation time with configuration A $t_{\textrm{obs-A}}$ in the time interval [5,~10]~hr, and that of configuration C $t_{\textrm{obs-C}}$ in the time interval [1,~{3}]~hr, (iii) the frequency bandwidth such that the ratio between the highest and the lowest frequencies $\rho_{\textrm{freq}}$ is in the range [1,~2], and (iv) the number of frequencies $n_{\textrm{freq}}$ between 1 and 4. Under these considerations, Fourier sampling patterns are created using the MeqTrees software \citep{Noordam2010}. Additionally, a random rotation and flagging of up to $20 \%$ of their data points are applied. The total number of points in the resulting Fourier sampling patterns ranges from $2\times 10^5$ to $2\times 10^6$.

The NUFFT measurement operators $\Phib$ are built from the generated sampling patterns, assuming a fixed ratio between the spatial Fourier bandwidth of the ground-truth images and the spatial bandwidth of the Fourier sampling that is set to 1.5. To enhance the effective resolution of the modeled visibilities, the measurement operators incorporate the Briggs weighting scheme \citep{briggs1995}. The scheme is standard in RI imaging as it enables an adjustable trade-off between resolution and sensitivity. The weights are generated using the WSClean software with the Briggs parameter set to 0~\citep{offringa2014,offringa2017}.

\begin{figure}
 \centering
 \setlength\tabcolsep{1pt} 
 \begin{tabular}{cccc}
 \includegraphics[width=0.24\linewidth]{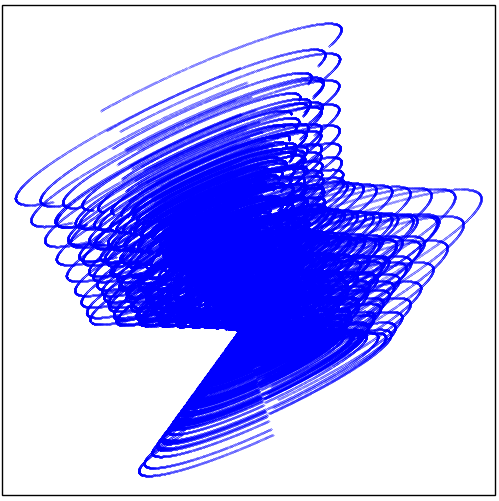} &
 \includegraphics[width=0.24\linewidth]{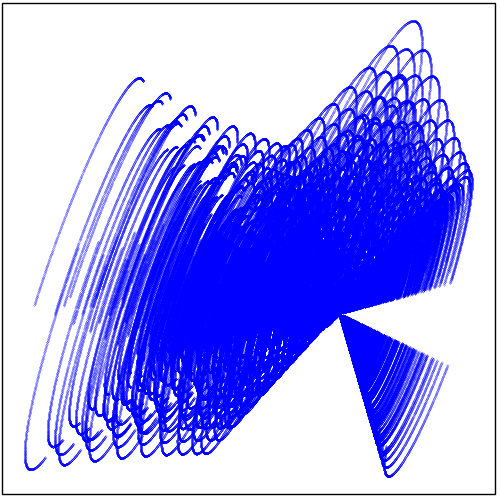} &
 \includegraphics[width=0.24\linewidth]{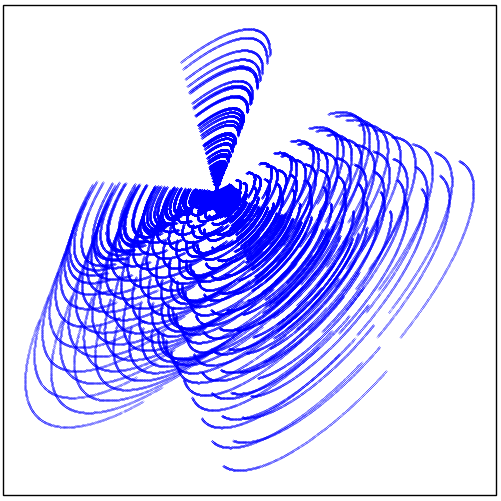} &
 \includegraphics[width=0.24\linewidth]{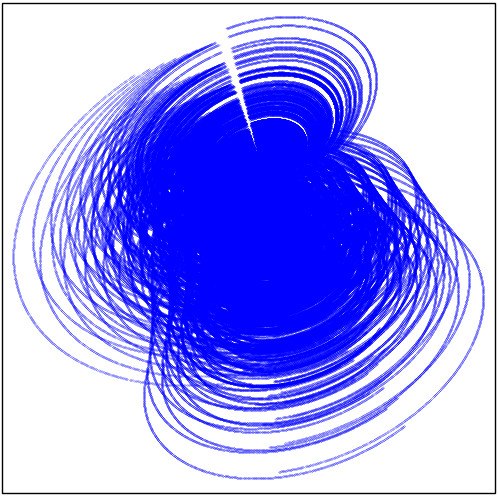}\\
 \includegraphics[width=0.24\linewidth]{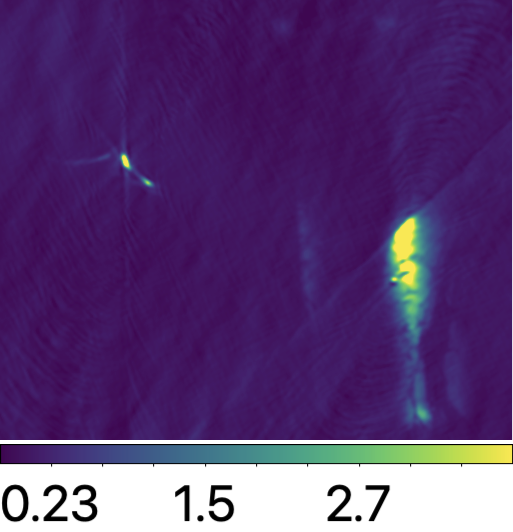} &
 \includegraphics[width=0.24\linewidth]{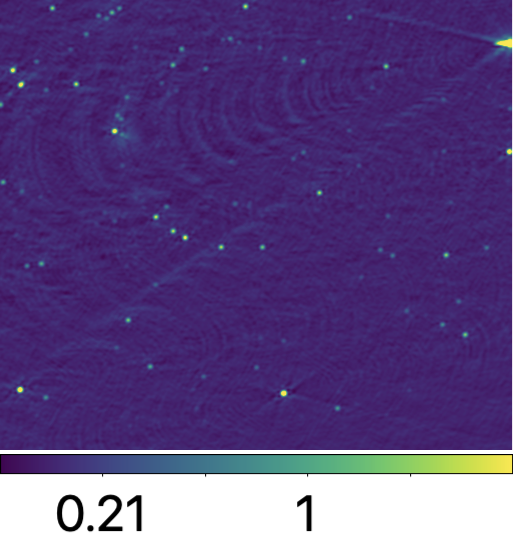} &
 \includegraphics[width=0.24\linewidth]{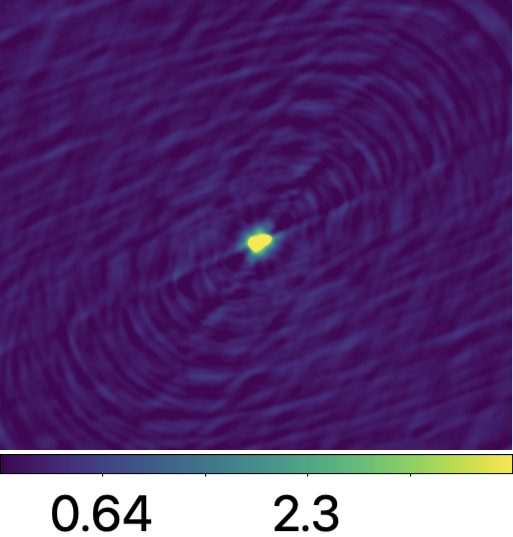} &
 \includegraphics[width=0.24\linewidth]{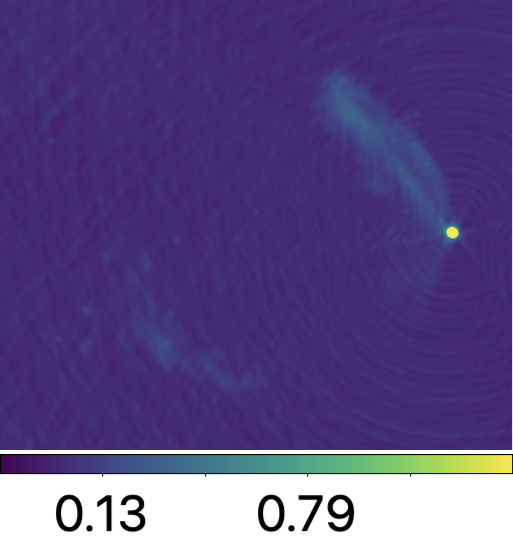}
 \end{tabular}
 \caption{A selection of simulated 2D Fourier sampling patterns with the antenna configuration of the VLA (top row) and resulting dirty images (linear scale; bottom row). The depicted dirty images correspond to the ground-truth images shown in Figure~\ref{fig:train_val}.
 \vspace*{-0.37cm}
 } 
 \label{fig:dirty_uv}
\end{figure}

For the ground-truth images, their dynamic range $\rho_{\textrm{DR}}$ is increased using the pixel-wise exponentiation transform described in \citet{terris2022} and is varied in the interval $[10^3,~ 5\times 10^5]$. More precisely, $\rho_{\textrm{DR}}$ is uniformly randomised in the logarithmic scale such that $\log_{10}(\rho_{\textrm{DR}})\in [3,~5.69]$. Using the data model \eqref{eq:observation}, modeled visibilities are corrupted with additive random Gaussian noise, with a standard deviation $\tau$ set in an adaptive manner following a stipulation proposed in \citet{terris2022} to ensure that the measurement noise adapts to the dynamic range of the radio image of interest. More precisely, for each ground truth image with a given dynamic range $\rho_{\textrm{DR}}$, $\tau$ is fixed such that:
\begin{equation}
{\tau = \eta_{\textrm{Briggs}}~(1/\rho_{\textrm{DR}}) \sqrt{2\|\textnormal{Re}\{{\Phib}^{\dagger}{\Phib}\}\|_S}},
\end{equation}
where $||\cdot||_S$ denotes the spectral norm of its argument operator, and $\eta_{\textrm{Briggs}}>0$ is a correction factor accounting for the Briggs weights, which reduces to 1 otherwise \citep{wilber23a}. The dirty images are then obtained via back-projection of the RI data as per \eqref{eq:backprojection}. The resulting images are of size $N=512\times 512$, same as the ground-truth images, with a pixel-size aligning with the super-resolution factor set to 1.5. Examples of simulated VLA sampling patterns and dirty images are provided in Figure~\ref{fig:dirty_uv}. 

Training and validation datasets are composed of pairs of high-dynamic range ground-truth images and associated dirty images. The training dataset incorporates ground-truth images generated from the sourced medical and optical astronomy images. The validation dataset incorporates ground-truth images obtained from the sourced radio astronomy images. The dichotomy in the nature of the images in both datasets aims to ensure the generalizability of the trained DNN series. The training and validation datasets consist of 20000 and 250 pairs of ground-truth images and associated dirty images, respectively. 

\subsection{Implementation} 
We train R2D2 and two realizations of R3D3 which are distinguished by the number of layers in their R2D2-Net components such that $J\in \{3,6\}$. Hereafter, we refer to these R3D3 realizations as $\textrm{R3D3}^{\textrm{3L}}$ and $\textrm{R3D3}^{\textrm{6L}}$. Similarly, we refer to their corresponding R2D2-Net realizations as $\textrm{R2D2-Net}^{\textrm{3L}}$ and $\textrm{R2D2-Net}^{\textrm{6L}}$.

In the sequential training of the different DNN series, we consider a pruning procedure which progressively reduces the size of the training dataset, initially set to $L=20000$ image pairs. It relies on a data fidelity-based criterion to decide on the convergence of the inverse problem associated with each training image pair $(\xb^\star_l,{\xb_{\textrm d}}_l)$ from the initial training dataset. More specifically, at any given iteration $i>1$, if the updated residual dirty image ${\rb^{(i)}_l}$ reaches the image-domain noise level by satisfying the condition $\| \rb_l^{(i)} \|^2_2 \leq \| \bb_l \|^2_2$, the pair $(\xb^\star_l,{\xb_{\textrm{d}}}_l)$ is discarded from the training dataset of subsequent DNNs in the series. The validation dataset, typically significantly smaller in size, remains unchanged throughout the training process.

The dataset pruning procedure can accelerate the training process and potentially reduce its computational cost, especially the cost incurred for the update of the residual dirty images. Furthermore, by eliminating solved inverse problems, the training of subsequent DNNs becomes focused on relevant inverse problems. The procedure can therefore improve the imaging precision of the DNN series. Interestingly, it also informs an additional stopping criterion for the training sequence, specifically, when the training dataset is reduced down to an inappropriate size for efficient DNN training. 

Training was conducted on Cirrus, a UK Tier 2 high-performance computing (HPC) service, equipped with both CPU and GPU compute nodes. The CPU nodes are composed of dual Intel 18-core Xeon E5-2695 processors with 256~GB of memory each. The GPU nodes are composed of two 20-core Intel Xeon Gold 6148 processors, four NVIDIA Tesla V100-SXM2-16GB GPUs, and 384~GB of DRAM memory. Computation of the dirty images and updates of the residual dirty images were run on the CPU nodes. DNNs' training relied on the PyTorch library in Python \citep{paszke2019pytorch}, and was performed on the GPU nodes. The learning rate was set to $10^{-4}$. Due to GPU memory constraints, a batch size of 4 and 1 was used for R2D2 and R3D3 realizations, respectively. Note that given the manageable scale of both image and data sizes of the training datasets, the computation of the residual dirty images can be efficiently performed on GPUs.

\begin{figure}
 \centering
 \includegraphics[width = 0.35\textwidth]{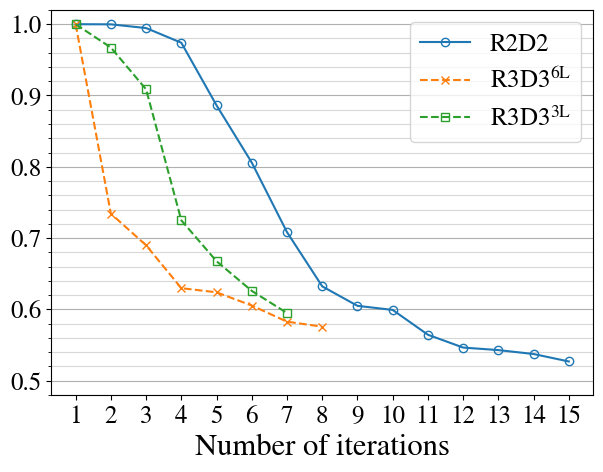}
 \caption{Results of the dataset pruning procedure. Evolution of the size of the training dataset shown as a fraction of the size of the initial training dataset, throughout the iterations of R2D2 and both R3D3 realizations ($\textrm{R3D3}^{\textrm{3L}}$,~$\textrm{R3D3}^{\textrm{6L}}$).}
 \label{fig:prun_table512}
 \vspace*{-0.3cm}
\end{figure}

R2D2 training concluded at iteration $I=15$. For R3D3 realizations, training of $\textrm{R3D3}^{\textrm{3L}}$ and $\textrm{R3D3}^{\textrm{6L}}$ concluded at iterations $I=7$ and $I=8$, respectively. The VLA-trained R2D2 and R3D3 DNN series are available in the dataset \citet{r2d2dnns}. Figure~\ref{fig:prun_table512} shows the evolution of the training dataset size during the training of R2D2 and R3D3 realizations ($\textrm{R3D3}^{\textrm{3L}}$, $\textrm{R3D3}^{\textrm{6L}}$). For R2D2, the reduction of the training dataset was triggered from the third iteration. For $\textrm{R3D3}^{\textrm{3L}}$ and $\textrm{R3D3}^{\textrm{6L}}$, the procedure was activated immediately after their first iteration, with up to 25\% of the training dataset discarded with the deeper network component $\textrm{R2D2-Net}^{\textrm{6L}}$. This suggests the ability of data model-informed DNN architectures to accelerate the data fidelity of the image estimates, which aligns with the smaller number of terms required in their associated DNN series. At least 40\% of the initial training dataset was discarded when training the last network component of all DNN series, suggesting the efficiency of the procedure. In principle, more network components can be trained with the remaining training datasets. However, in our current setting, the sequential training of all DNN series terminated based on the evaluation metrics of the validation dataset (see Section~\ref{subsec:DNN-series-training}).

\subsection{Computational cost}
The computational cost of building the different DNN series in terms of CPU core time (CPU~hr) incurred for the computation of the residual dirty images, and GPU time (GPU~hr) incurred for the training of the DNNs, is provided in Table~\ref{table:training_cost}. The computational cost of the first network component of their series is also provided. On the one hand, R2D2 required approximately 80\% less GPU~hr than both realizations of R3D3, owing to the simpler architecture of its network components. On the other hand, it demanded about twice as many CPU~hr for updating the residual dirty images, since it is trained with almost twice as many network components in its series. Interestingly, for all DNN series, 30 to 60\% of the GPU~hr cost was incurred in the training of their first network component. Their subsequent network components benefited from efficient initialization of their learnable parameters using their preceding trained component, and a reduced training dataset thanks to the data fidelity-driven pruning procedure. Both realizations of R3D3 required similar resources in both GPU~hr and CPU~hr, suggesting a balance between the depth of their network components and the required number of iterations in their series. 
\begin{table}
\hspace{-2cm}
\resizebox{1.27\columnwidth}{!}{%
\begin{tabular}{lccccccc}\toprule
 & {$I$} & {$Q (\times10^6)$} & {$n_{\textrm{epochs}}$} & $n_{\textrm{gpu}}$ & $n_{\textrm{cpu}}$ & GPU~hr & CPU~hr \\\midrule
U-Net &1 & 31 & 264& 4 & 6 & 82 & 336 \\
$\textrm{R2D2-Net}^{\textrm{3L}}$ &1 & 93 & 405& 12 & 6 & 873 & 336 \\
$\textrm{R2D2-Net}^{\textrm{6L}}$ &1 & 186 & 192& 12 & 6 & 414 & 336\\
R2D2  &15 & 31 & 398& 4 & 6 & 160 & 4757 \\
$\textrm{R3D3}^{\textrm{3L}}$ &7 & 93 & 605& 12 & 6 & 1291 & 2165 \\
$\textrm{R3D3}^{\textrm{6L}}$ &8 & 186 & 591& 12 & 6 & 1276 & 2244\\\bottomrule
\end{tabular}
}
\caption {Training computational details of R2D2, R3D3 realizations ($\textrm{R3D3}^{\textrm{3L}}$,~$\textrm{R3D3}^{\textrm{6L}}$), and the end-to-end DNNs corresponding to the first components in their series, U-Net and R2D2-Net ($\textrm{R2D2-Net}^{\textrm{3L}}$, $\textrm{R2D2-Net}^{\textrm{6L}}$), respectively. Results are reported in terms of: the number of iterations~($I$), the number of learnable parameters of their network components~($Q$), the cumulative number of epochs~($n_{\textrm{epochs}}$), the number of CPU cores~($n_{\textrm{cpu}}$) deployed for generating the dirty images and updating the residual dirty images, and the number of GPUs~($n_{\textrm{gpu}}$), deployed for DNN training and updating the image estimates. The training computational cost is provided in GPU~hr and CPU~hr.} 
\label{table:training_cost}
\end{table}

\section{Simulations and results} \label{sec:simulations-results} 
In this section, we study the performance of the VLA-trained incarnations of the R2D2 algorithm in terms of the imaging precision. We present the RI imaging algorithms used as a benchmark including their parameter choice and the adopted evaluation metrics. In our analysis, two experimental setups are considered. The first is generic, where the image and observation settings are fully randomized. The second is focused on exploring different regimes of key parameters of the image and observation settings. The computational efficiency of the proposed algorithm is studied using its implementations in Python and MATLAB.
\subsection{Benchmark algorithms \& parameter choice}
The performance of the R2D2 algorithm is evaluated in comparison with the sparsity-based algorithm from optimization theory uSARA \citep{repetti2020forward,terris2022}, the PnP algorithm AIRI \citep{terris2022,terris2023plug}, and multiscale CLEAN \citep{cornwell2008multiscale} in the WSClean software \citep{offringa2014,offringa2017}. All algorithms formally feature free parameters, whose values should be fixed following an appropriate procedure. Firstly, uSARA features a parameter that controls the trade-off between its handcrafted regularization function and data fidelity. AIRI features a parameter controlling the appropriate noise level of its underpinning DNN denoiser. The choice of uSARA and AIRI parameter is automated using noise-driven heuristics \citep{terris2022,dabbech2022,wilber23a}. Adjustments from the heuristic values can sometimes be required for optimized results. Secondly, in WSClean, multiscale CLEAN parameters are set according to nominal values, often requiring adjustments for improved results. In particular, the parameters controlling the cleaning depth can require tuning as they directly impact the data fidelity of the CLEAN reconstruction. For the proposed simulations, optimized values of the uSARA and AIRI heuristics, and the CLEANing depth were studied and fixed to the same value for batches of simulated inverse problems (assuming an automated/pipeline mode reconstruction) rather than for each reconstruction specifically (see Section~\ref{subsec:Experiment-generic-settings}). Finally, as discussed in Section~\ref{subsec:DNN-series-training}, we note that the R2D2 algorithm is here seen as parameter-free with the number of its network components determined at the training stage {(the number of iterations is set to the number of trained DNNs).} This represents an interesting advantage over the benchmark in the perspective of the deployment of the imaging algorithms in automated mode.

\subsection{Implementation}
For the benchmark algorithms, we use multiscale CLEAN implemented in C++ as part of the WSClean software, and uSARA and AIRI MATLAB implementations in the {BASPLib}\footnote{https://basp-group.github.io/BASPLib/} code library. The R2D2 algorithm (in its two incarnations) comes in two distinct implementations in MATLAB and Python. The former shares the NUFFT implementation of AIRI and uSARA, and the latter uses a fast PyTorch-specific implementation \citep{muckley20}. Although both NUFFT implementations are based on the works of \citet{Fessler2003}, the Python version uses a table-based interpolation, whereas the MATLAB version uses a more precise sparse matrix interpolation. The end-to-end U-Net and R2D2-Net, representing the first iterations of R2D2 and R3D3, benefit from both MATLAB and Python implementations.

All algorithms were executed on Cirrus, utilizing similar resources where relevant. Given the relatively small image and data dimensions considered in this work, minimal resources were allocated. CLEAN and uSARA were run using 1 CPU core each. AIRI, R2D2, and R3D3, in their MATLAB implementation, utilized 1 CPU core for data-fidelity specific operations and 1 GPU for the DNN inference. The MATLAB implementation of the end-to-end U-Net and R2D2-Net used a similar allocation, dedicating the CPU core for the computation of the dirty image. For the Python implementation of R2D2 and R3D3 (respectively, U-Net and R2D2-Net), two resource allocation settings are supported: (i) a hybrid setting that is consistent with their MATLAB version which used 1 CPU core for the update of the residual dirty images (respectively, computation of the dirty image) and 1 GPU for the DNN inference, (ii) and a full GPU implementation which used 1 GPU.

\subsection{Evaluation metrics}
A qualitative and quantitative evaluation of the reconstruction quality obtained by the different algorithms is provided through the visual examination of the image estimates, and the signal-to-noise ratio metric, both in linear scale (SNR) and logarithmic scale (logSNR). 
More precisely, considering a ground truth image $\xb^\star$ and an image estimate $\widehat{\xb}$, the SNR metric reads:
\begin{equation}
\textrm{SNR}(\widehat{\xb}, {\xb^{\star}}) = 20\textrm{log}_{10}(\| {\xb^{\star}}\|_2 / \|{\xb^{\star}} - \widehat{\xb}\|_2).
\end{equation}
The logSNR metric, a variation of SNR, is introduced to assess the reconstruction quality with emphasis on faint structure. For this purpose, we consider the following logarithmic mapping of the images of interest, parametrized by $a>0$:
\begin{equation}
\textrm{rlog}(\xb) = x_{\textrm{max}}\textrm{log}_{a}({\frac{a}{x_{\textrm{max}}}}~ \xb+\textbf{1}),
\end{equation}
where $x_{\textrm{max}}$ is the peak pixel value of the image $\xb$ and $\textbf{1} \in \eR^N$ is a vector of values equal to 1. Setting the parameter $a$ to $\rho_{\textrm{DR}}$, the dynamic range of the ground truth image, the logSNR metric reads:
\begin{equation}
 \textrm{log}\textrm{SNR}(\widehat{\xb}, {\xb^{\star}}) = \textrm{SNR}(\textrm{rlog}(\widehat{\xb}), \textrm{rlog}({\xb^{\star}})).
\end{equation}
We also examine image-domain data fidelity obtained by all algorithms. Considering a dirty image $\xb_{\textrm{d}}$ and an estimated residual dirty image $\widehat\rb$, the metric denoted by $\overline{\sigma}_{\textrm{res.}}$ reads:
\begin{equation} 
\overline{\sigma}_{\textrm{res.}}(\widehat{\rb}, \xb_{\textrm{d}})=\|\widehat{\rb}\|_2/\|\xb_{\textrm{d}}\|_2.
\label{eq:data_fidelity}
\end{equation}
Additionally, we provide a qualitative assessment via the visual inspection of the estimated residual dirty images. 

The computational performance of the different algorithms is analyzed using the total number of iterations $I$, the total computational time $t_{\textrm{tot.}}$, as well as the average computational time per iteration for the data-fidelity step $t_{\textrm{dat.}}$ and the regularization step $t_{\textrm{reg.}}$. Since all algorithms used a single CPU core and/or a single GPU, the computational time is reported in seconds.

\subsection {Experiment in generic image \& data settings}
\label{subsec:Experiment-generic-settings}

The experimental setup is generic in that all parameters characterizing the ground-truth images and the observations were uniformly randomized, as per the considerations of the training (see Section~\ref{sec:Training-approach} for details).
Test datasets were created using four real radio images: the giant radio galaxies 3C353 (NRAO Archives) and Messier~106 \citep{shimwell2022}, and the radio galaxy clusters {Abell~2034} and PSZ2~G165.68+44.01 \citep{botteon2022}. 
From each radio image, 50 ground-truth images of size $N =512 \times 512$ with varying dynamic range were generated following the pre-processing procedure used for the training datasets. Different Fourier sampling patterns were used to simulate the RI data associated with the ground-truth images. The resulting test dataset comprises 200 inverse problems.

The parameter choice of the benchmark algorithms is as follows. The uSARA parameter was set to two times the heuristic value. The AIRI parameter was fixed to the heuristic value for all RI data, except for those simulated from the radio image of 3C353, where 3 times the heuristic value was used. CLEAN parameters were set to ensure \textit{deep} cleaning with minimal compromise on the algorithm stability\footnote{WSClean command:~\tt{wsclean -niter 2000000 -weight briggs 0 -j 1 -multiscale -mgain 0.8 -gain 0.1 -auto-threshold 0.5 -auto-mask 1.5 -padding 2}}. In particular, both auto-masking and threshold parameters were set to 1.5 and 0.5 times the estimate of the noise level, respectively. 

Numerical reconstruction results of the different algorithms are summarized in Table~\ref{table:results}. The reported values correspond to the average across the 200 inverse problems. SNR and logSNR metrics showcase the overall superior performance of R2D2, and both realizations of R3D3 ($\textrm{R3D3}^{\textrm{3L}}$,~$\textrm{R3D3}^{\textrm{6L}}$), with more than 3~dB improvement over AIRI and uSARA. CLEAN results indicate sub-optimal performance with more than 10~dB lower values than uSARA and AIRI. {This is somewhat expected since CLEAN reconstruction is limited in both resolution (due to the smoothing with the CLEAN beam) and dynamic range (due to the addition of the residual dirty image). Furthermore, it does not inherently enforce the nonnegativity constraint in the context of intensity imaging.} Focusing on R2D2 and R3D3, they both showcase comparable SNR and logSNR values, with incremental improvement achieved by the latter. Specific to R3D3, its deeper realization $\textrm{R3D3}^{\textrm{6L}}$ provides {an incremental} improvement over $\textrm{R3D3}^{\textrm{3L}}$ in SNR {of about 0.2~dB on average}, and exhibits similar results in logSNR. {These results showcase the consistency of the reconstruction quality of R3D3 across variations of the depth of its network components, and more generally the consistency of the imaging precision of the R2D2 algorithm in its two incarnations.}  

For the end-to-end DNNs, the fully data-driven U-Net provides a sub-optimal reconstruction, where both SNR and logSNR are at least 10~dB lower than R2D2, demonstrating the interest of the series structure underpinning the proposed deep-learning approach to enable high imaging precision. Interestingly, both realizations of R2D2-Net ($\textrm{R2D2-Net}^{\textrm{3L}}$, $\textrm{R2D2-Net}^{\textrm{6L}}$) are able to achieve good reconstruction. On the one hand, $\textrm{R2D2-Net}^{\textrm{3L}}$ delivers a 2~dB higher SNR value than both AIRI and uSARA, and a lower logSNR value by the same amount. On the other hand, $\textrm{R2D2-Net}^{\textrm{6L}}$ outperforms both AIRI and uSARA by more than 2~dB in both SNR and logSNR, showcasing the benefit of increasing the number of layers in the data model-informed R2D2-Net to enhance the dynamic range of the estimated images. Although the deeper realization of R2D2-Net reduces the gap in the reconstruction quality with R3D3, the latter provides 1~dB improvement in logSNR, confirming the interest of the series structure of the proposed algorithm. 

Numerical analysis of the image-domain data fidelity obtained by the imaging algorithms (see Table \ref{table:results}) indicates that CLEAN delivers the highest data fidelity, enabled by the thorough cleaning functionality in WSClean. Both realizations of R3D3, uSARA and AIRI obtain low $\overline{\sigma}_{\textrm{res.}}$ values that are comparable to CLEAN. R2D2, however, provides a value nearly twice as high as R3D3 realizations, suggesting lower data fidelity. These results showcase the importance of the architecture of the network components of the R2D2 algorithm to enhance the data fidelity of its reconstruction.

The Python and MATLAB implementations of R2D2 and R3D3 yield consistent results, with marginal difference in logSNR values induced by their different NUFFT implementations. We note that sparse matrix interpolation NUFFT implementation is also available in Python. 
\begin{rotatetable*}
\centerwidetable
\begin{deluxetable*}{lcccccccccc}
\tablecaption{Results of the experiment in generic image and data settings obtained by the imaging algorithms. {Reconstruction quality metrics are SNR, logSNR, and ${\overline{\sigma}}_{\textrm{res.}}$. Computational details are presented in terms of: the total number of iterations ($I$), the total reconstruction time~($t_{\textrm{tot.}}$), the average time per iteration of the data fidelity step~($t_{\textrm{dat.}}$) and the regularization step ($t_{\textrm{reg.}}$), and the allocated resources in GPUs~($n_{\textrm{gpu}}$) and CPU cores~($n_{\textrm{cpu}}$). Reported values are computed as averages over 200 inverse problems. Information on the programming languages underlying the algorithms implementations is provided.} \label{table:results}}
\tablecolumns{11}
\tablehead{
\colhead{} &
\colhead{SNR}&
\colhead{logSNR} & 
\colhead{$\overline{\sigma}_{\textrm{res.}}$} & 
\colhead{$I$} & 
\colhead{{$t_{\textrm{tot.}}$}} &
\colhead{{$t_{\textrm{dat.}}$}} &
\colhead{{$t_{\textrm{reg.}}$}} &
\colhead{$n_{\textrm{cpu}}$} &
\colhead{$n_{\textrm{gpu}}$} &
\colhead{Programming}\\
\colhead{} &
\colhead{$\pm$~std~(dB)}&
\colhead{$\pm$~std~(dB)} & 
\colhead{$\pm$~std~($\times$1E-4)} & 
\colhead{$\pm~$~std} & 
\colhead{$\pm$~std~(s)} &
\colhead{$\pm$~std~(s)} &
\colhead{$\pm$~std~(s)} &
\colhead{} &
\colhead{} &
\colhead{language} 
}
\startdata
{CLEAN} & \lowq{13.6$\pm$3.6} & \lowq{10.3$\pm$3.5} & \highq{5.1$\pm$5.2} & \highq{9$\pm$1} & \highq{65.9$\pm$18.8} & {3.8$\pm$1.1} & 3.8$\pm$1.0 & 1 & - & C++ \\
uSARA & \highq{30.8$\pm$1.9} & 
\highq{21.9$\pm$3.3} & \highq{6.5$\pm$8.2} & \lowq{1103$\pm$373} & \lowq{4184.2$\pm$1548.9} & 1.4$\pm$0.7 & 2.3$\pm$1.1 & 1 & - & MATLAB\\
{AIRI} & \highq{31.3$\pm$2.3} & \highq{21.9$\pm$4.4} & \highq{6.4$\pm$8.0} & \lowq{5000$\pm$0.0} & \lowq{3478.8$\pm$1531.4} & 0.66$\pm$0.3 & 0.03$\pm$0.2 & 1 & 1 & MATLAB\\
\midrule
\multirow{2}{*}{U-Net} & \lowq{20.5$\pm$2.7} & \lowq{6.6$\pm$3.3} & \lowq{777.5$\pm$467.4} & \highq{1} & \highq{2.2$\pm$0.6} & - & 2.2$\pm$0.6 & - & 1 & MATLAB\\
\cline{2-11}
& \lowq{20.5$\pm$2.7} & \lowq{6.6$\pm$3.3} & \lowq{777.6$\pm$467.4} & \highq{1} & \highq{1.1$\pm$0.1} & - & 1.1$\pm$0.1 & - & 1 & Python\\
\midrule
\multirow{2}{*}{R2D2-Ne$\textrm{t}^{\textrm{3L}}$} & \highq{32.6$\pm$1.5} & \highq{19.6$\pm$5.5} & \highq{21.5$\pm$13.8 } & \highq{1} & \highq{2.7$\pm$0.4 }& - & 2.7$\pm$0.4 & - & 1 & MATLAB\\
\cline{2-11}
& \highq{32.6$\pm$1.5} & \highq{19.6$\pm$5.4} & \highq{21.5$\pm$13.8} & \highq{1} & \highq{1.1$\pm$0.1} & - & 1.1$\pm$0.1 & - & 1 & Python\\
\midrule
\multirow{2}{*}{R2D2-Ne$\textrm{t}^{\textrm{6L}}$} & \highq{33.7$\pm$1.7} & \highq{24.0$\pm$4.7} & \highq{9.3$\pm$6.9} & \highq{1} & \highq{3.8$\pm$1.5} & - & 3.8$\pm$1.5 & - & 1 & MATLAB\\
\cline{2-11}
& \highq{33.7$\pm$1.7} & \highq{24.0$\pm$4.7} & \highq{9.3$\pm$7.0} & \highq{1} & \highq{1.1$\pm$0.1} & - & 1.1$\pm$0.1 & - & 1 & Python\\
\midrule
\multirow{3}{*}{R2D2} & \highq{33.7$\pm$1.5} & \highq{25.1$\pm$4.9} & \highq{13.5$\pm$46.9} & \highq{15} & \highq{12.2$\pm$2.8} & 0.4$\pm$0.2 & 0.2$\pm$0.07 & 1 & 1 & MATLAB\\
\cline{2-11}
& \multirow{2}{*}{\highq{33.7$\pm$1.5}} & \multirow{2}{*}{\highq{25.0$\pm$4.9}} & \multirow{2}{*}{\highq{13.5$\pm$46.9}}& \multirow{2}{*}{\highq{15}}& \highq{18.6$\pm$5.9} & 1.1$\pm$0.4 & 0.1$\pm$0.1 & 1 & 1 & Python\\
&&&&& \highq{2.9$\pm$0.3} & 0.05$\pm$0.01 & 0.1$\pm$0.2 & - & 1 & Python\\
\midrule
\multirow{3}{*}{$\textrm{R3D3}^{\textrm{3L}}$} & \highq{33.8$\pm$1.4} & \highq{25.3$\pm$4.6} & \highq{7.6$\pm$7.6} & \highq{7} & \highq{9.4$\pm$1.5} & 0.5$\pm$0.2 & 0.5$\pm$0.1 & 1 & 1 & MATLAB\\
 \cline{2-11}
& \multirow{2}{*}{\highq{33.8$\pm$1.4}} & \multirow{2}{*}{\highq{25.3$\pm$4.6}} & \multirow{2}{*}{\highq{7.6$\pm$7.6}} & \multirow{2}{*}{\highq{7}} & \highq{9.8$\pm$4.3} & 1.1$\pm$0.4 & 0.2$\pm$0.3 & 1 & 1& Python\\
&&&&& \highq{1.9$\pm$0.5} & 0.05$\pm$0.02 & 0.2$\pm$0.3 & - & 1& Python\\
\midrule
\multirow{3}{*}{$\textrm{R3D3}^{\textrm{6L}}$} & \highq{34.0$\pm$1.6} & \highq{25.3$\pm$4.7} & \highq{7.9$\pm$7.8} & \highq{8} & \highq{15.2$\pm$2.1} & 0.5$\pm$0.2 & 1.3$\pm$0.6 & 1 & 1 & MATLAB\\
 \cline{2-11}
& \multirow{2}{*}{\highq{34.0$\pm$1.6}}& \multirow{2}{*}{\highq{25.3$\pm$4.7}} & \multirow{2}{*}{\highq{7.9$\pm$7.8}} & \multirow{2}{*}{\highq{8}}& \highq{11.3$\pm$3.9 }& 1.1$\pm$0.4 & 0.2$\pm$0.3& 1 & 1& Python\\
&&&&& \highq{2.2$\pm$0.3} & 0.05$\pm$0.02 & 0.2$\pm$0.3 & - & 1& Python\\
\bottomrule
\enddata
\tablenotetext{}{{\textbf{Note}. For enhanced readability, a color-coding is considered to categorize the performance of the imaging algorithms: high in green, sub-optimal in red.} Specific to CLEAN, the reported number of iterations corresponds to the number of ``major cycles'' reached at convergence. Two inverse problems from the test dataset diverged and are therefore excluded from the reported results. These instances of instability in the CLEAN implementation could potentially stem from the \textit{deep} cleaning.}
\end{deluxetable*}
\end{rotatetable*}

\noindent
We note that sparse matrix interpolation NUFFT implementation is also available in Python. This was validated on the same experiments and confirmed to provide identical results to the MATLAB version. It is however slower than its table-based interpolation counterpart.

\subsection {Experiments in specific image \& data settings}
\label{subsec:Experiment-specific-settings}
Considering the four radio images used in the previous experiment, we designed four experiments (I-IV) to assess the performance of both incarnations of the R2D2 algorithm (i) in contrasting regimes of the dynamic range of the sought radio images, and (ii) in varying observation settings, in terms of the total observation time and the bandwidth of the frequency channels (under the assumption of radio emission with flat spectra). The simulation parameter choice in the four experiments is listed in Table~\ref{table:testset_config}. In all experiments, the dynamic range of the ground-truth images is fixed. Therefore, 1 ground truth image per radio image was generated. For each experiment, we simulated 25 RI datasets with varying Fourier sampling patterns per ground truth image, resulting in a total of 100 inverse problems. Imaging with the benchmark algorithms, in particular their parameter choice, followed the same considerations of the generic experiment.

Numerical results of Experiments~I--IV are provided in Figure~\ref{fig:metrics_plots}, comprising graphs of the evolution of the SNR, logSNR, and $\overline{\sigma}_{\textrm{res.}}$ metrics across the iterations of R2D2 and R3D3. In these graphs, the numerical values at the first iteration of R2D2, $\textrm{R3D3}^{\textrm{3L}}$, and $\textrm{R3D3}^{\textrm{6L}}$ correspond to the results of the end-to-end DNNs, U-Net, $\textrm{R2D2-Net}^{\textrm{3L}}$, and $\textrm{R2D2-Net}^{\textrm{6L}}$, respectively. Final results of CLEAN, AIRI, and uSARA are indicated with horizontal lines.

\begin{table}
 \centering
 \begin{tabular}{cccccc}
 \toprule
 {Identifier} & {$\rho_{\textrm{DR}}$} & {$t_{\textrm{obs-A}}$ (hr)} & {$t_{\textrm{obs-C}}$ (hr)} &{ ${n}_{\textrm{freq}}$} & {$\rho_{\textrm{freq}}$} \\
 \midrule
 I & $10^5$ & 5.5 & 1.1 & 1 & 1 \\
 II & {$5\times 10^3$} & 5.5 & 1.1 & 1 & 1 \\
 III & $10^5$ & 5.5 & 1.1 & 4 & 2 \\
 IV & $10^5$ & 9.0 & 2.7 & 1 & 1 \\
 \bottomrule
 \end{tabular}
\caption{Details of Experiments~I--IV in terms of: the value of the dynamic range of the ground-truth images ($\rho_{\textrm{DR}}$), the total duration of the observations with VLA configurations A and C $(t_{\textrm{obs-A}},t_{\textrm{obs-C}})$, the number of frequency channels ($n_{\textrm{freq}}$), and the ratio between the highest and lowest frequency channels ($\rho_{\textrm{freq}}$). With Experiment~I taken as the reference, the different settings were designed to study the impact of each parameter in two different regimes.}
 \label{table:testset_config}
 \vspace*{-0.3cm}
\end{table}

In Experiment~I, we consider a single frequency-channel observation, a relatively short observation time totaling 6.6 hr with both VLA configurations, and an extremely high-dynamic range regime with $\rho_{\textrm{DR}}=10^5$. This scenario will serve as a reference in our analysis. We first study the performance of R2D2 and R3D3 under different regimes of the dynamic range of the ground-truth images. To this aim, we consider Experiment~II, characterized by a relatively low-dynamic range regime with $\rho_{\textrm{DR}}=5\times 10^3$. In terms of SNR and logSNR, both $\textrm{R3D3}^{\textrm{3L}}$ and $\textrm{R3D3}^{\textrm{6L}}$ consistently outperform all imaging algorithms, showing a wider gap in both SNR and logSNR in the high-dynamic range regime. Although trailing behind AIRI in the low-dynamic range regime, R2D2 outperforms the benchmark algorithms in the high-dynamic range regime. Generally, both realizations of R3D3 converge faster than R2D2, with the SNR and logSNR metrics saturating more promptly in the low-dynamic range regime compared to the high-dynamic range regime.

We study the impact of multi-frequency acquisition and longer observation duration through Experiments~III and IV, both enabling more Fourier information of the sought images than Experiment~I, by combining wideband observations in the former, and increasing the combined total observation time to 11.7~hr in the latter. When compared to Experiment~I, both incarnations of the R2D2 algorithm obtain higher SNR and logSNR values, outperforming the benchmark algorithms. Moreover, their metrics saturate more promptly. These experiments showcase consistency in the results of the R2D2 algorithm.

Concerning data fidelity, in the low-dynamic range regime (Experiment~II), all algorithms obtain similar $\overline{\sigma}_{\textrm{res.}}$ values, suggesting comparable data fidelity. In the high-dynamic range regime (Experiments~I,~III--IV), AIRI and uSARA consistently exhibit the lowest $\overline{\sigma}_{\textrm{res.}}$ values, thus superior data fidelity, whereas CLEAN obtains 2 to 4 times higher values. {Both R3D3 realizations deliver similar results, which are 3 to 5 times higher than AIRI and uSARA. R2D2, however, delivers the highest $\overline{\sigma}_{\textrm{res.}}$ values among all algorithms, indicating the lowest data fidelity. Yet, when compared to R3D3, R2D2 results are only 20\% to 60\% higher.

\begin{figure*}
 \setlength{\tabcolsep}{1pt}
 \begin{tabular}{ccc}
 
 \includegraphics[width = 0.32\textwidth]{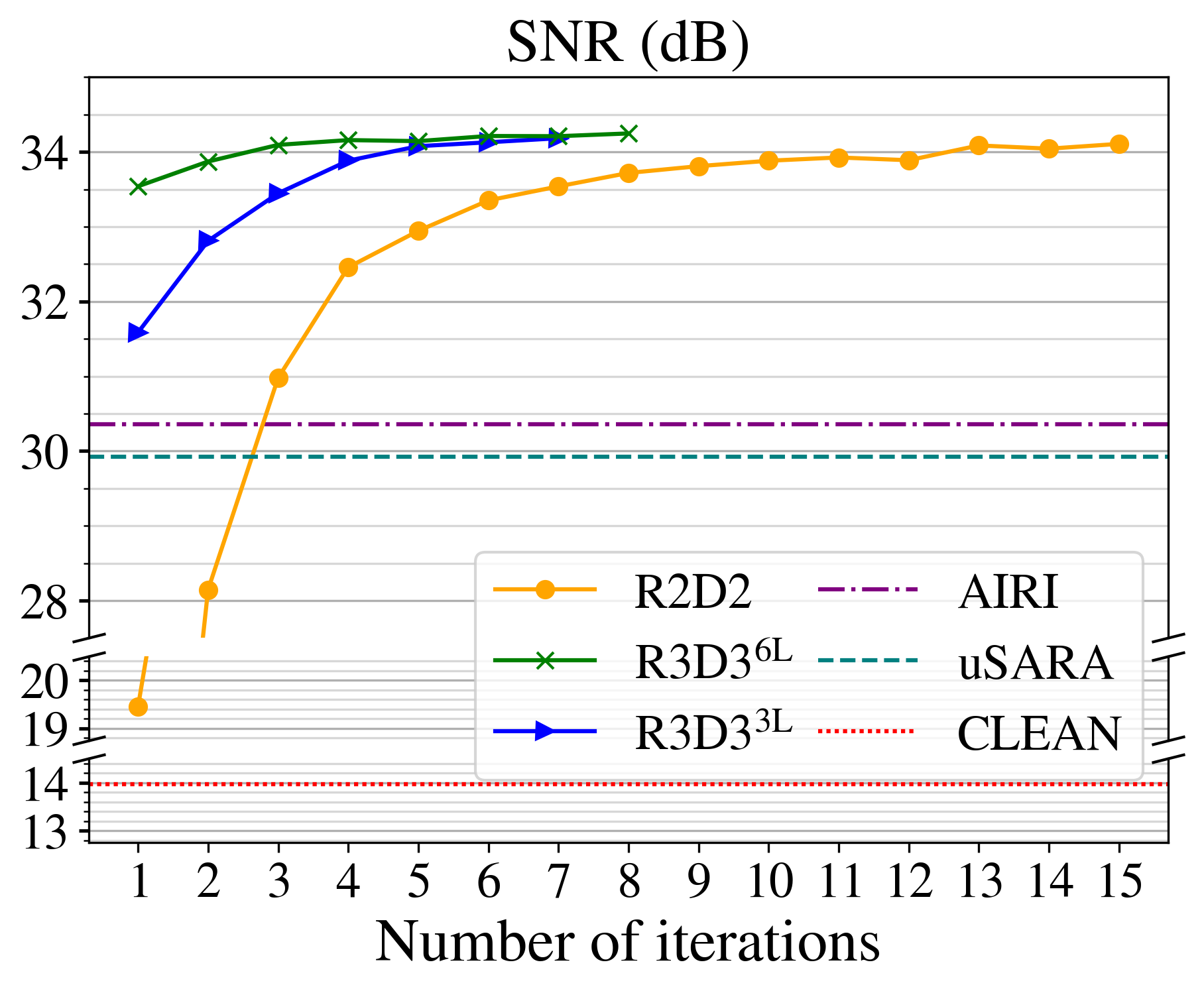}&
 \includegraphics[width = 0.32\textwidth]{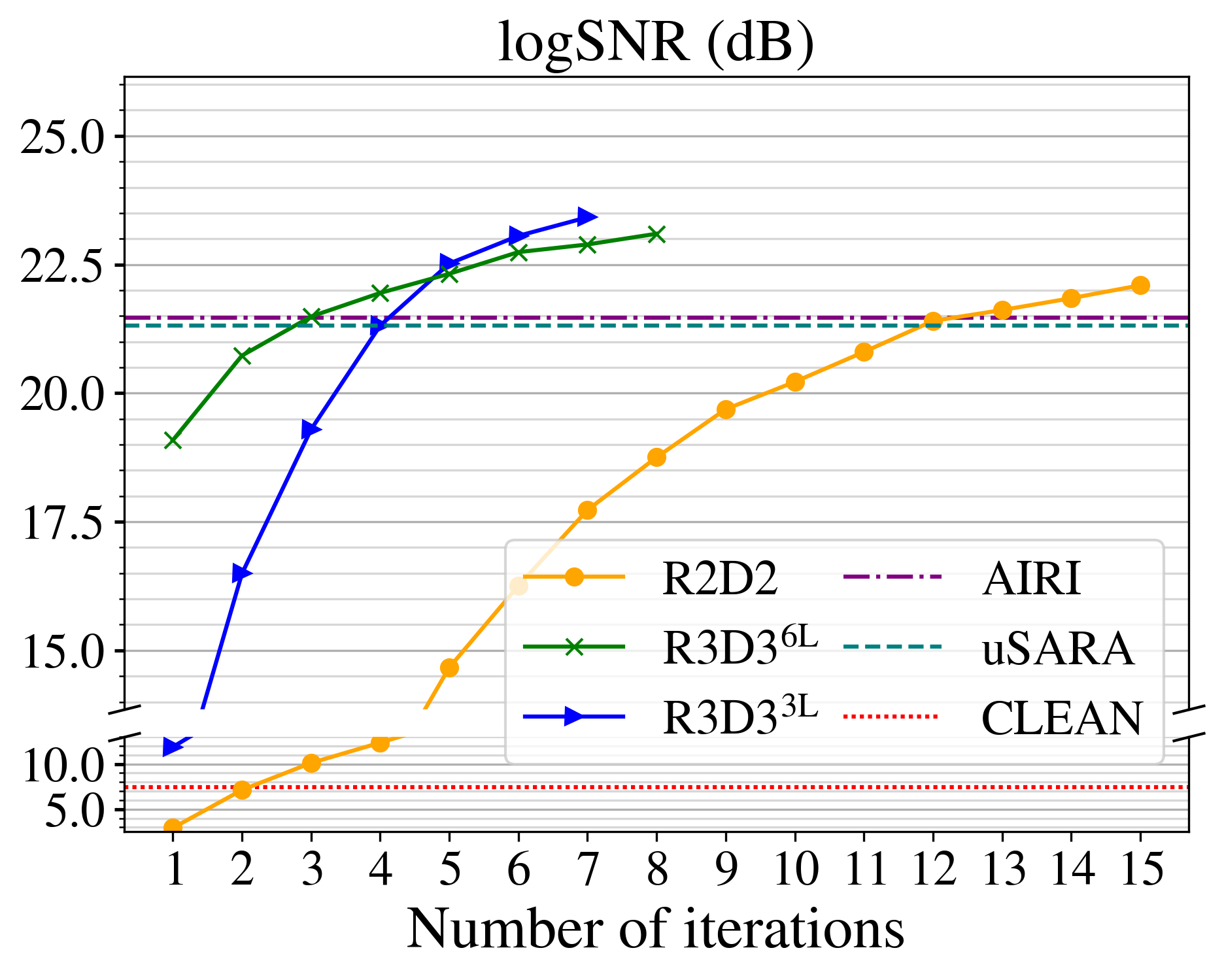}&
 \includegraphics[width = 0.32\textwidth]{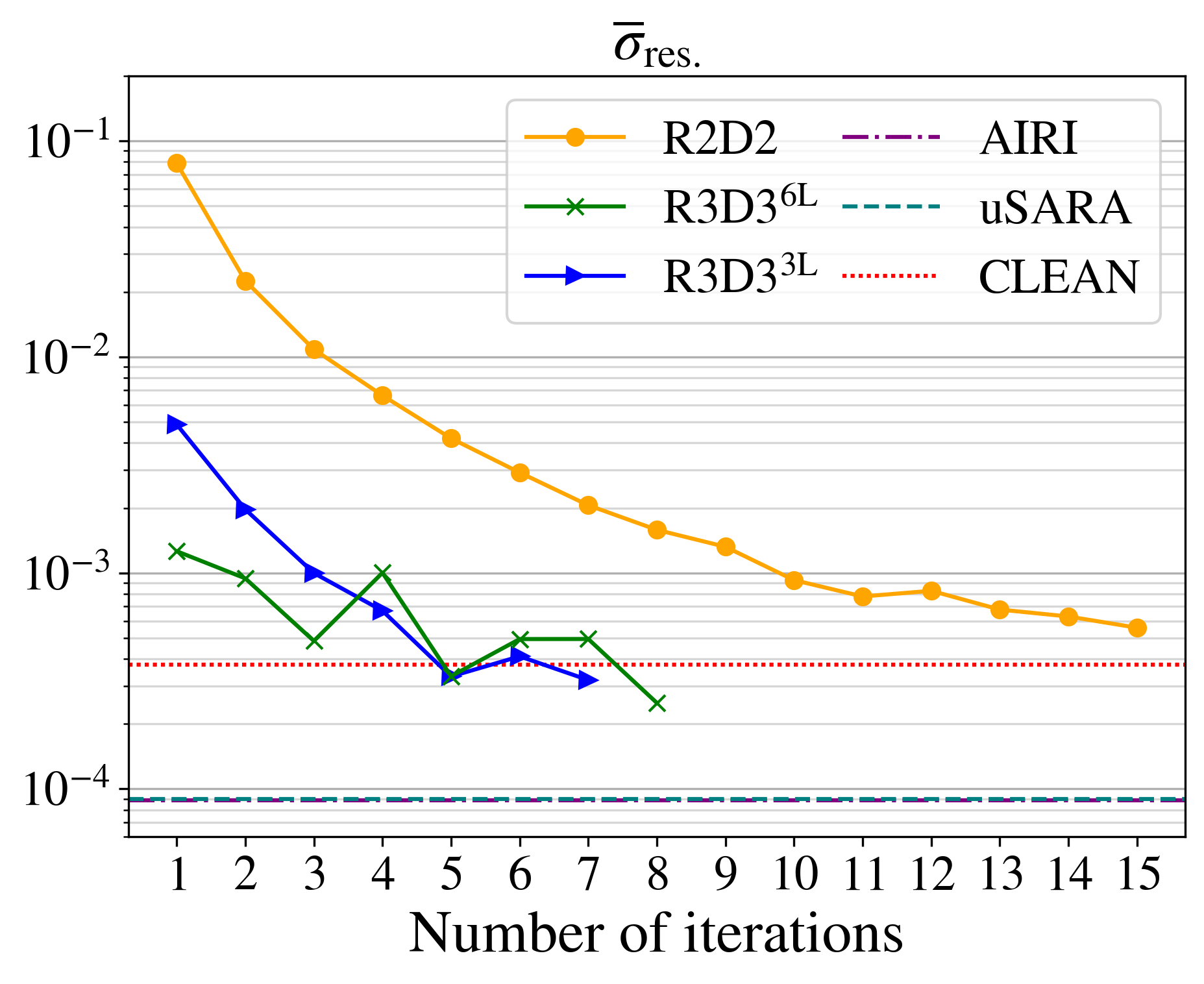}\\
 \multicolumn{3}{c}{Experiment~I:~$\rho_{\textrm{DR}}= 10^5$, $(t_{\textrm{obs-A}},t_{\textrm{obs-C}})=(5.5,1.1)$~hr, ${n}_{\textrm{freq}}=1$}\\
 \\
 
 \includegraphics[width = 0.32\textwidth]{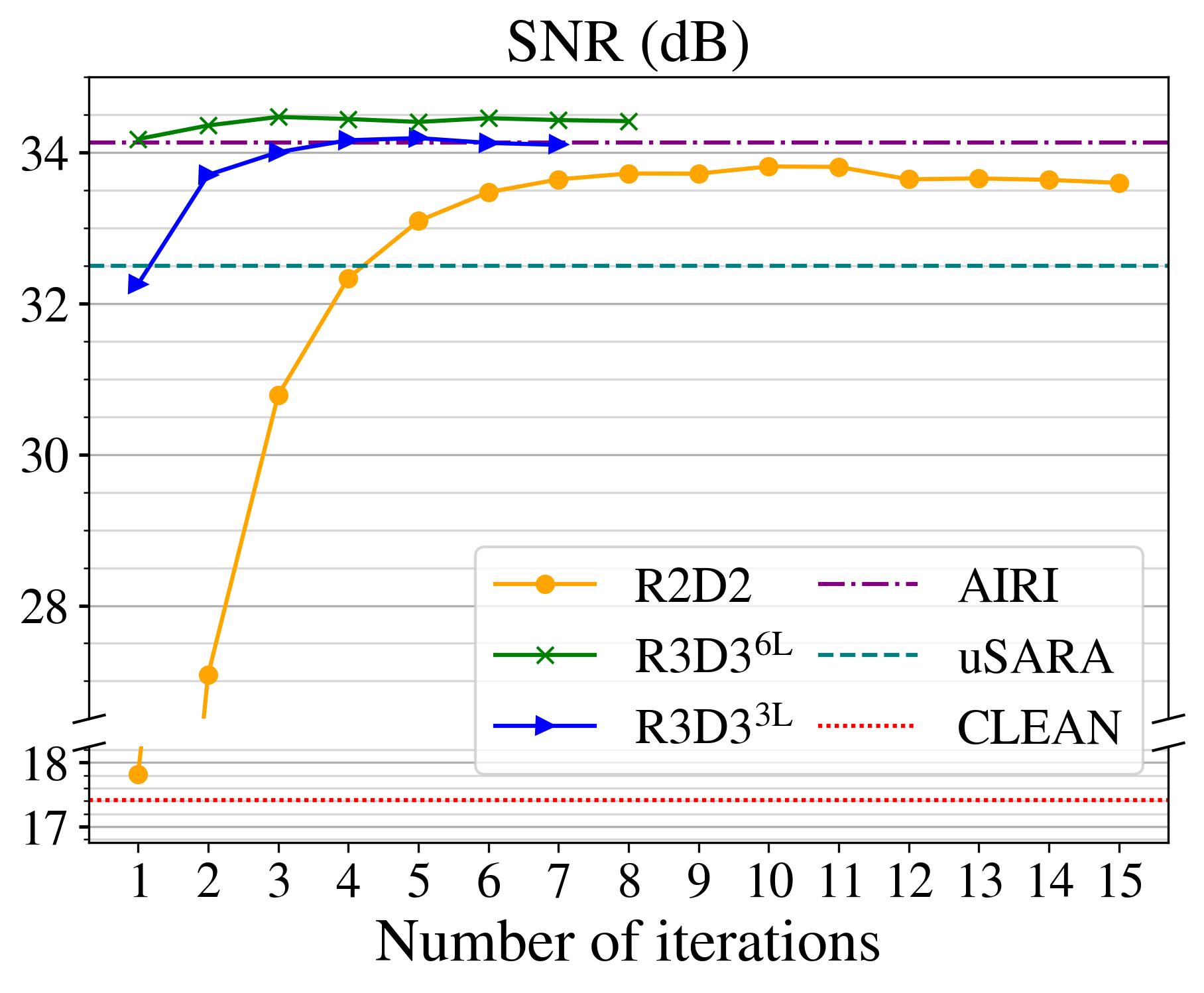}&
 \includegraphics[width = 0.32\textwidth]{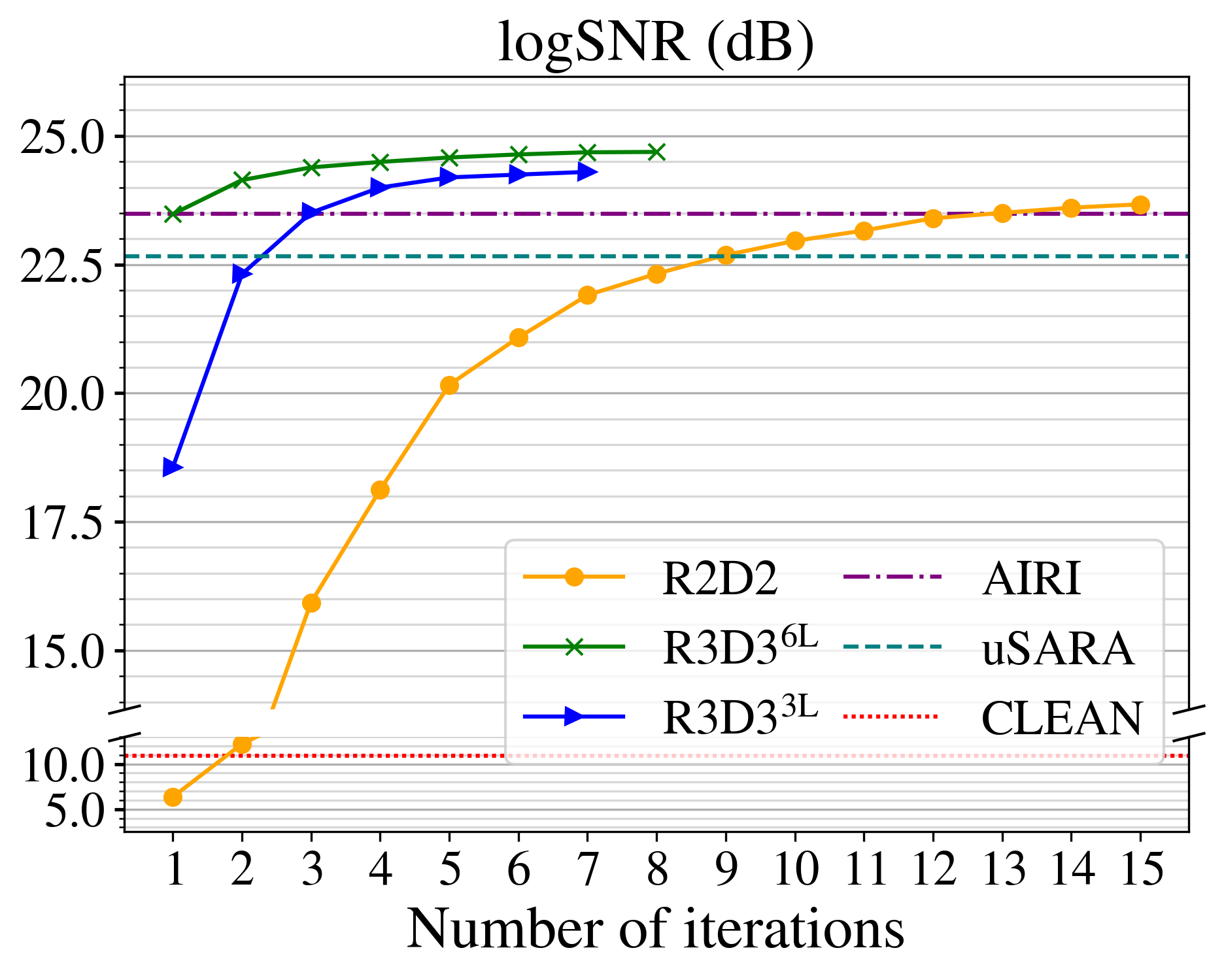}&
 \includegraphics[width = 0.32\textwidth]{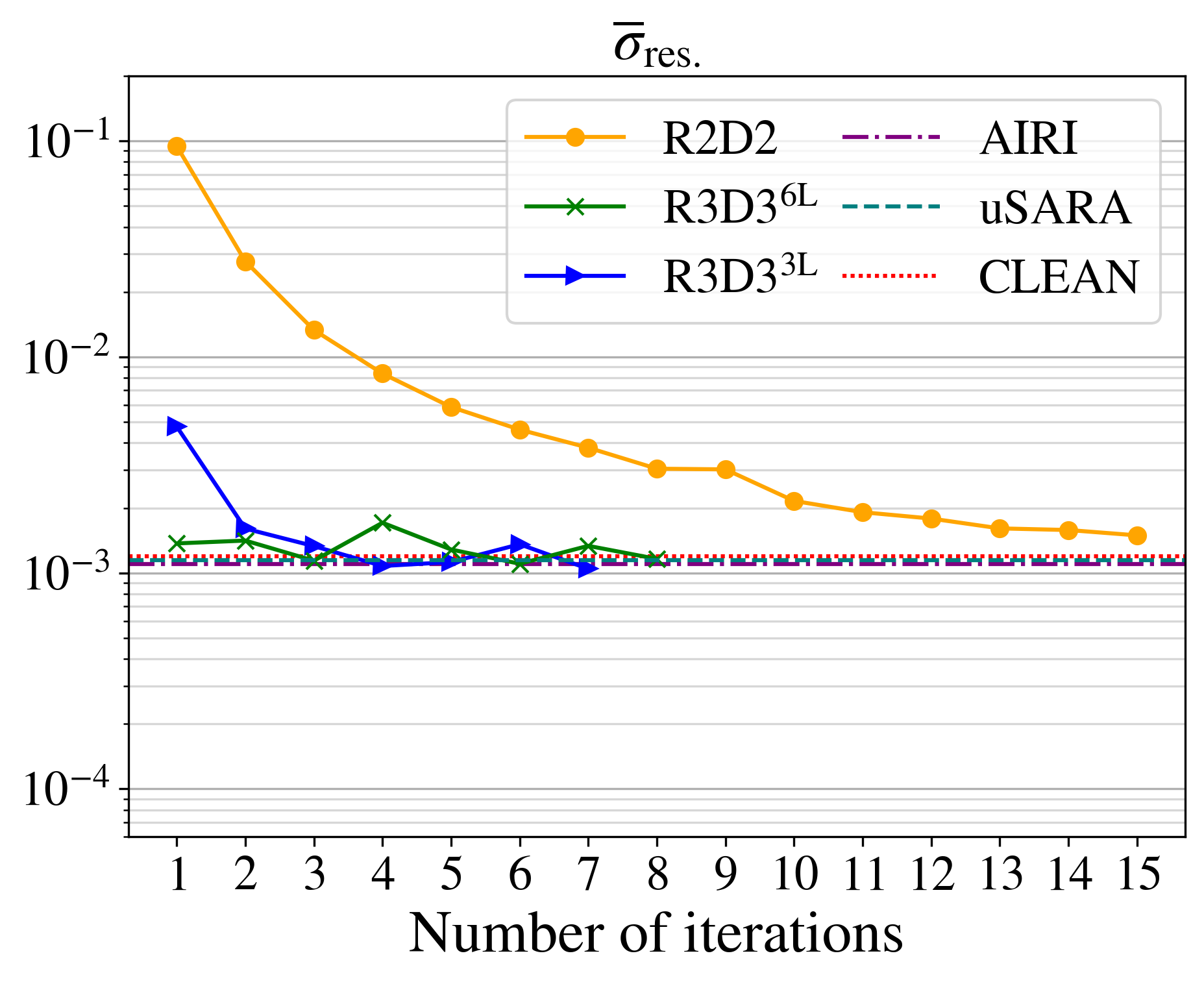}\\
 \multicolumn{3}{c}{Experiment~II:~$\rho_{\textrm{DR}} = 5\times 10^3$, $(t_{\textrm{obs-A}},t_{\textrm{obs-C}})=(5.5,1.1)$~hr, ${n}_{\textrm{freq}}=1$}\\
 \\

 \includegraphics[width = 0.32\textwidth]{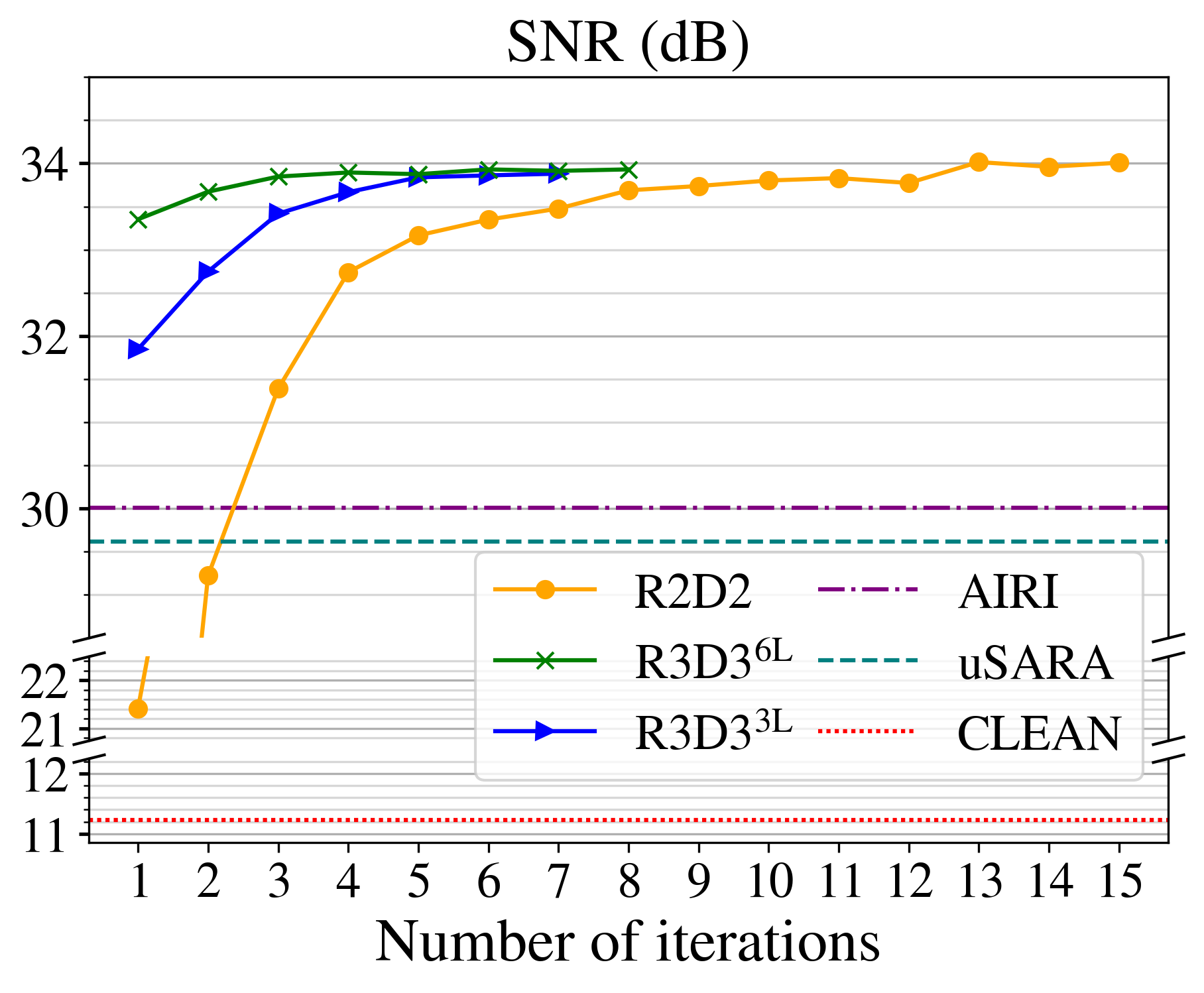}&
 \includegraphics[width = 0.32\textwidth]{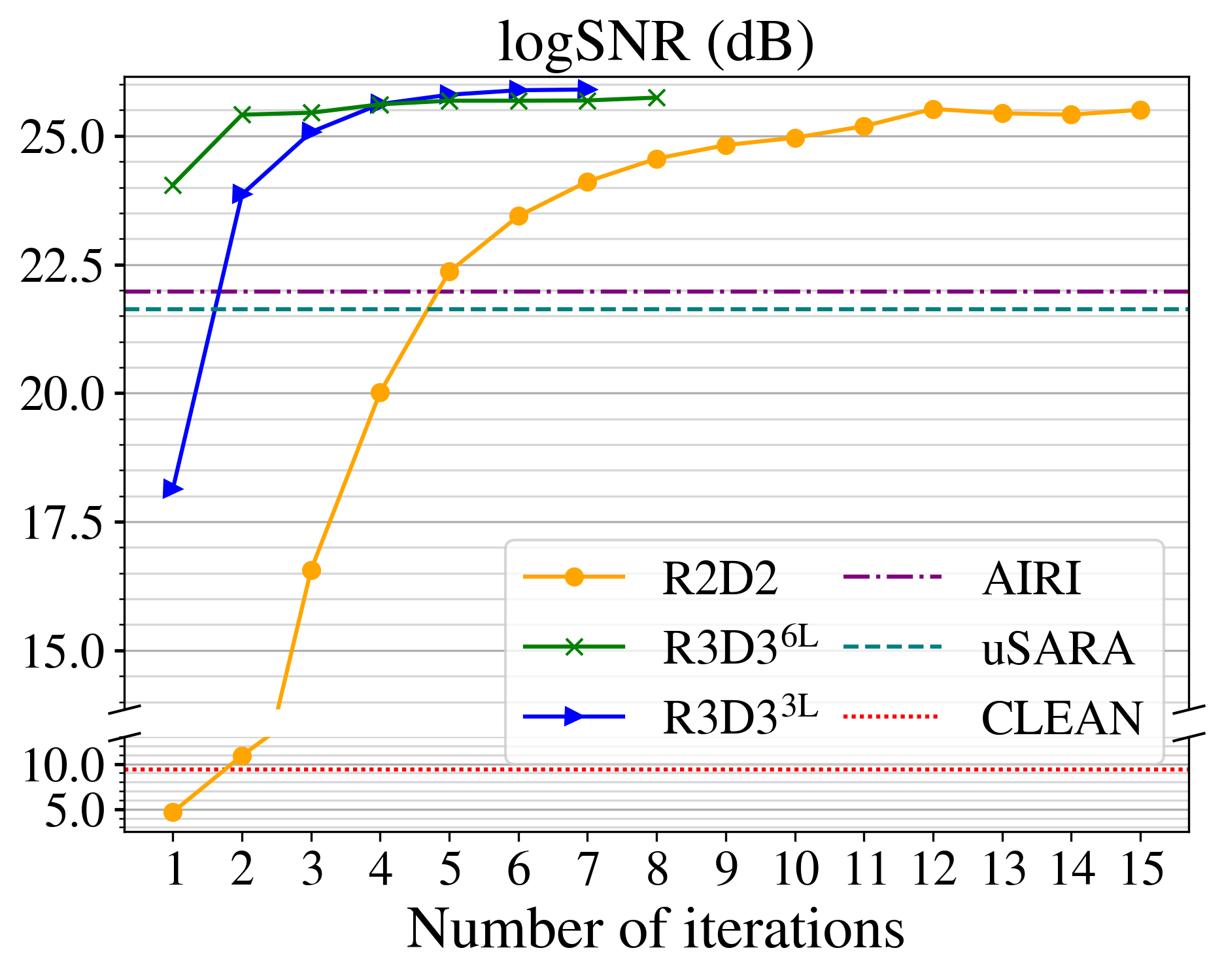} &
 \includegraphics[width = 0.32\textwidth]{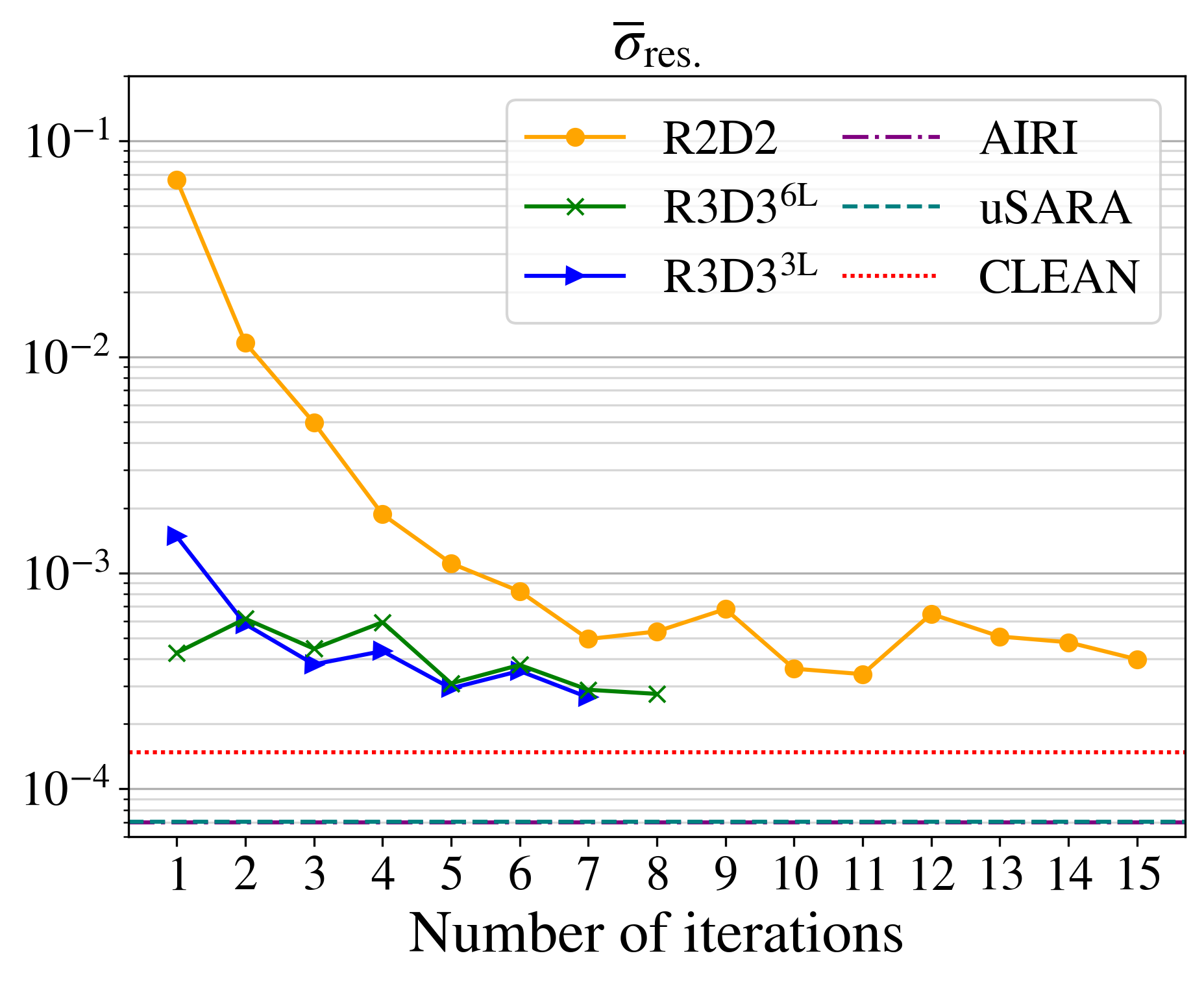}\\
 \multicolumn{3}{c}{Experiment~III:~$\rho_{\textrm{DR}}= 10^5$, $(t_{\textrm{obs-A}},t_{\textrm{obs-C}})=(5.5,1.1)$~hr, $\rho_{\textrm{freq}} =2$, ${n}_{\textrm{freq}}=4$} \\
 \\
 
 \includegraphics[width = 0.32\textwidth]{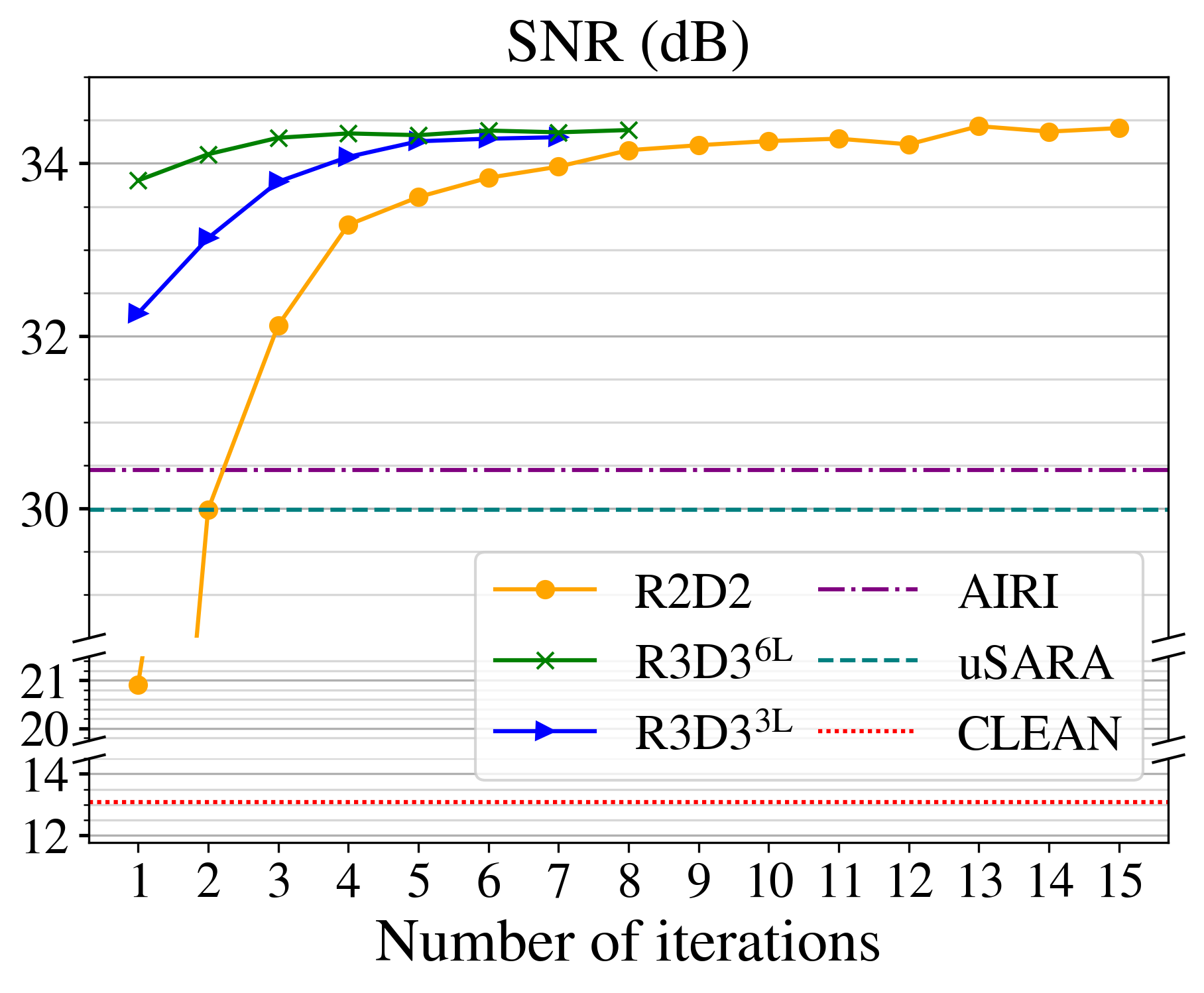}&
 \includegraphics[width = 0.32\textwidth]{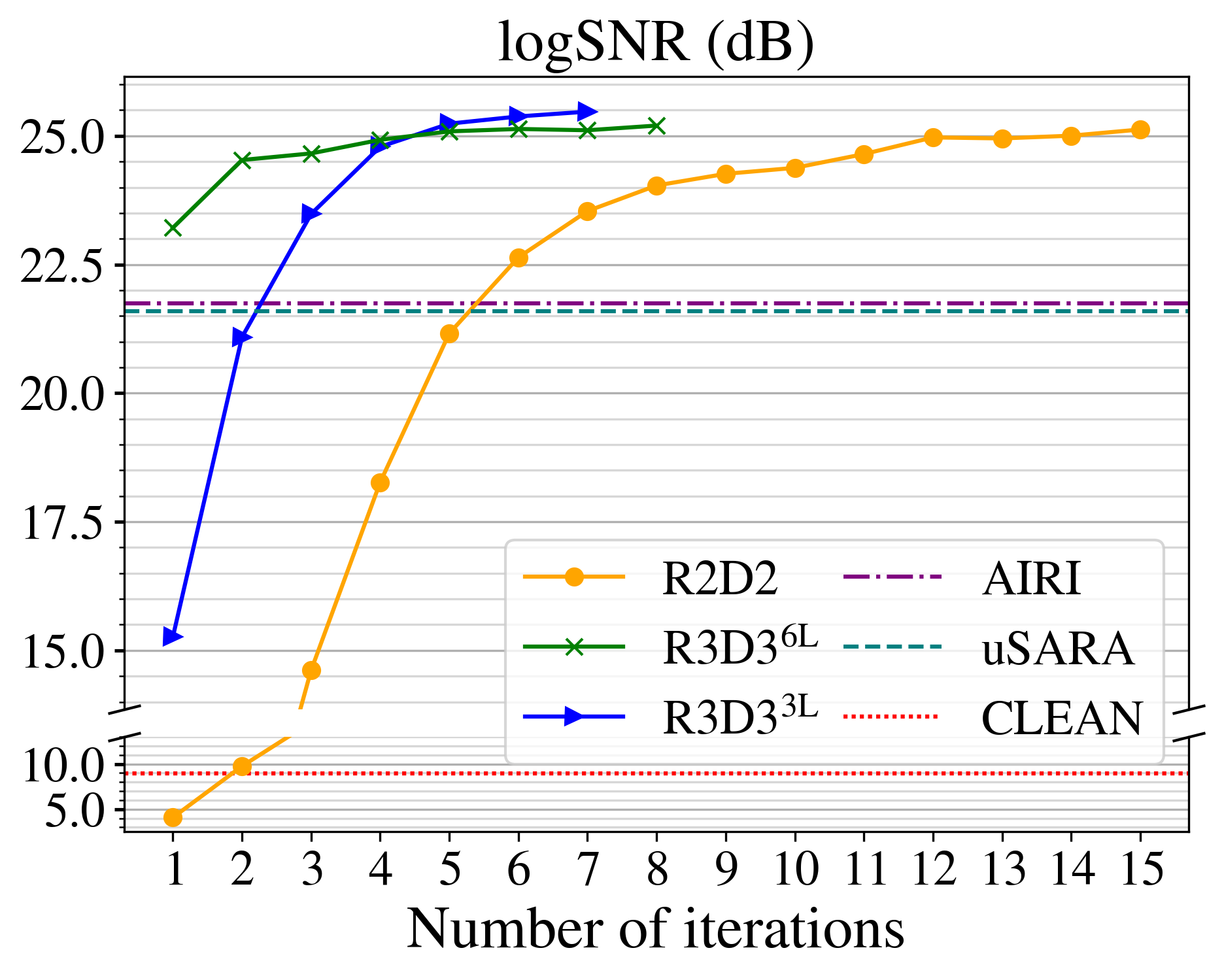}&
 \includegraphics[width = 0.32\textwidth]{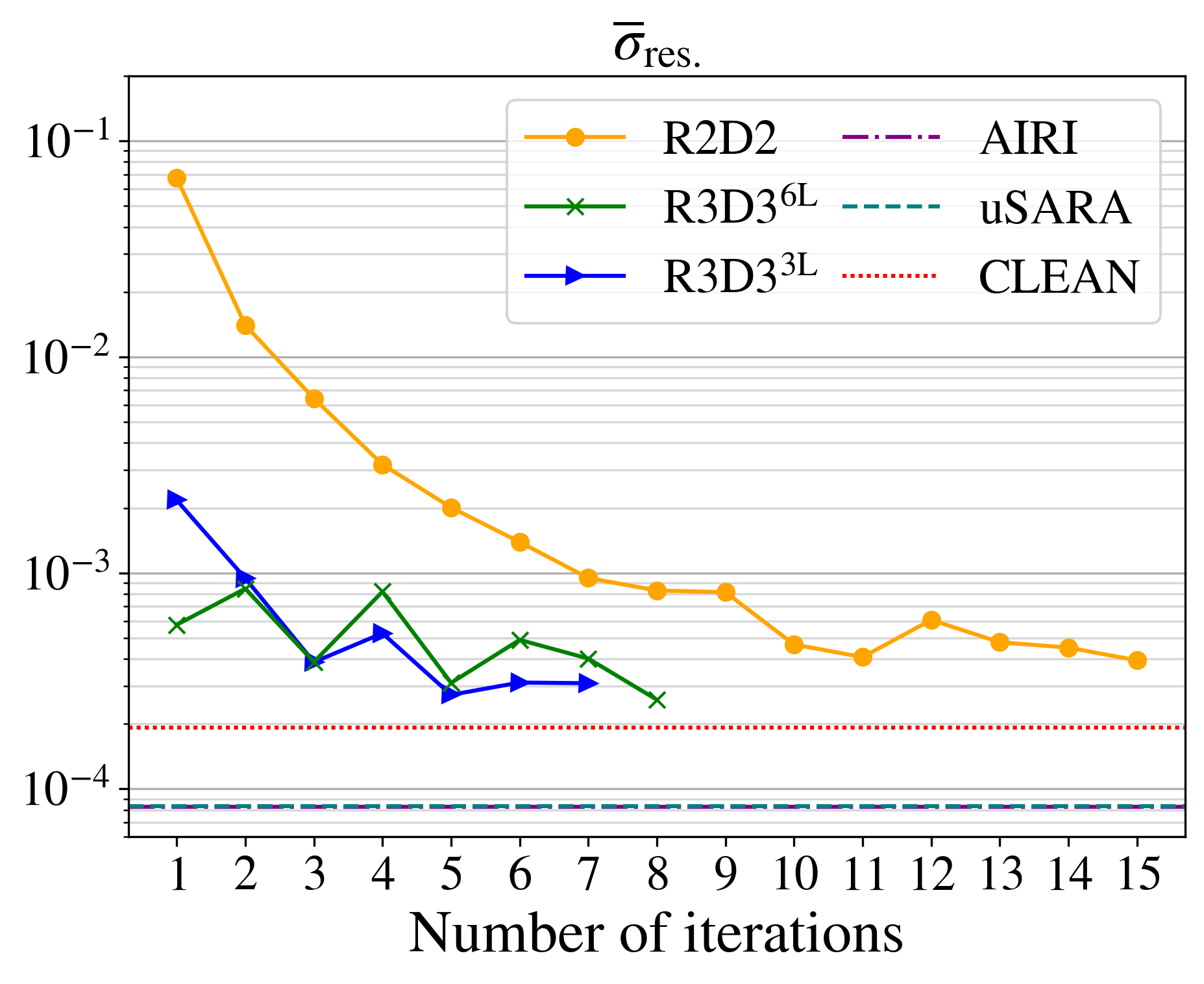}\\
 \multicolumn{3}{c}{Experiment~IV:~$\rho_{\textrm{DR}}= 10^5$, $(t_{\textrm{obs-A}},t_{\textrm{obs-C}})=(9,2.7)$~hr, ${n}_{\textrm{freq}}=1$}\\
 \end{tabular} 
\caption{Progress of the reconstruction quality (left column: SNR, middle column: logSNR) and data fidelity (right column: ${\overline{\sigma}}_{\textrm{res.}}$) across the iterations of R2D2 and both realizations of R3D3. Values of the different metrics, achieved at convergence by the benchmark algorithms uSARA, AIRI and CLEAN, are reported via horizontal lines. From top to bottom: results of Experiments~I-IV. In each scenario, the reported values are averages over a test dataset composed of 100 inverse problems.
}
\label{fig:metrics_plots}
\end{figure*}

Reconstructed images of selected test sets from Experiments~I--IV are displayed in Figures~\ref{fig:testset1}--\ref{fig:testset4}, respectively. The scrutinized radio images exhibit different morphology, from extended to compact and point-like structure. Since both realizations of R3D3 yield comparable results, we consider reconstructed images $\textrm{R3D3}^{\textrm{6L}}$ in this visual examination. One can observe that R2D2, R3D3, AIRI and uSARA reconstructions are comparable. CLEAN reconstruction, by design, provides a smoother depiction of the radio emission and a noise-like background. Focusing on the end-to-end DNNs, U-Net delivers an overly-smooth representation, {and in some cases, hallucination structure (e.g. the extended emission in the bottom right quadrant in the reconstruction of the cluster PSZ2 G165.68+44.0 in Figure~\ref{fig:testset3}, and the compact source near Messier 106 in Figure~\ref{fig:testset4}). Such artifacts are completely removed in the R2D2 reconstruction, highlighting its robustness, induced by its iterative nature.} R2D2-Net achieves a high-quality reconstruction, confirming the numerical analysis. {In contrast with U-Net, R2D2-Net does not exhibit hallucination structure, owing to its integrated data consistency layers.} The examination of the residual dirty images shows that R2D2 and R3D3 consistently present discernible structures around the pixel positions of the brightest emission in the radio images, particularly noticeable in the high-dynamic range regime (see bottom row of Figures~\ref{fig:testset2}--\ref{fig:testset4}). Both AIRI and uSARA achieve high data fidelity with noise-like residual dirty images. CLEAN obtains residual dirty images comparable to AIRI and uSARA, except for the selected test set from Experiment~I. Fine-tuning manually the cleaning depth for this specific test set could improve the results. These findings are in general agreement with the numerical analysis. 

\begin{figure*}
 \centering
 \setlength\tabcolsep{1pt} 
 \begin{tabular}{cccc}
 \includegraphics[width=0.23\linewidth]{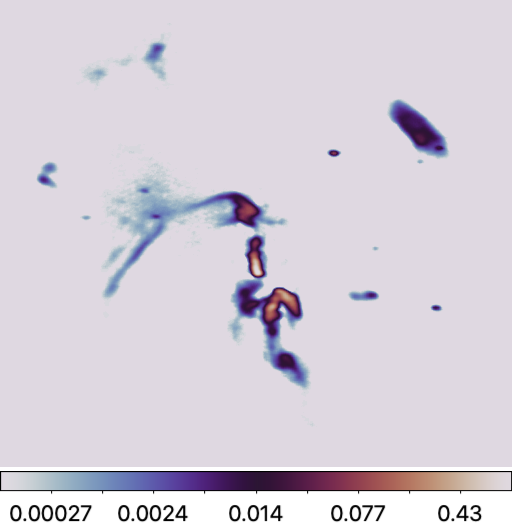} &
 \includegraphics[width=0.23\linewidth]{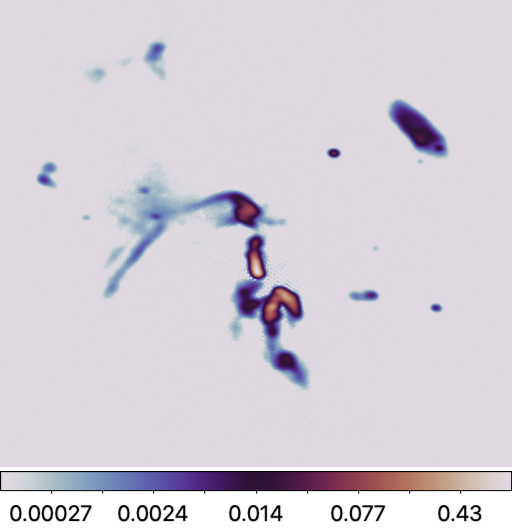} &
 \includegraphics[width=0.23\linewidth]{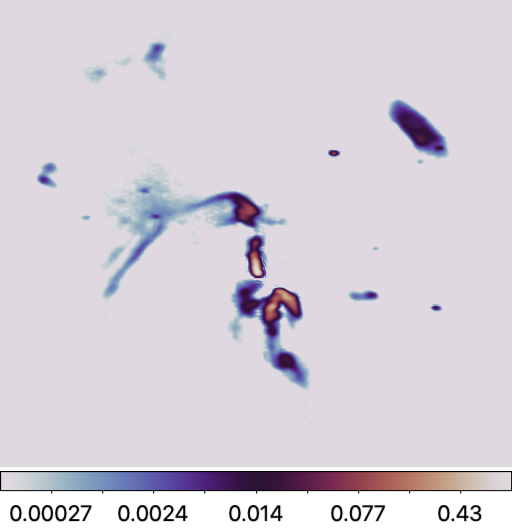} &
 \includegraphics[width=0.23\linewidth]{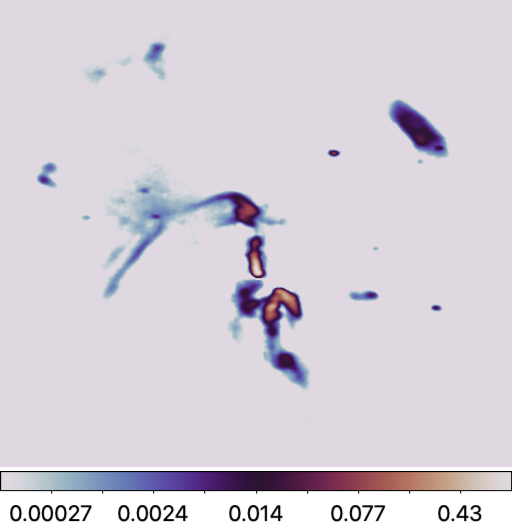}\\
 Ground truth $\xb^{\star}$ & CLEAN: (16.3, 8.3)~dB & uSARA: (31.9, {19.7})~dB & AIRI: (32.1, 19.{8})~dB \\

 \includegraphics[width=0.23\linewidth]{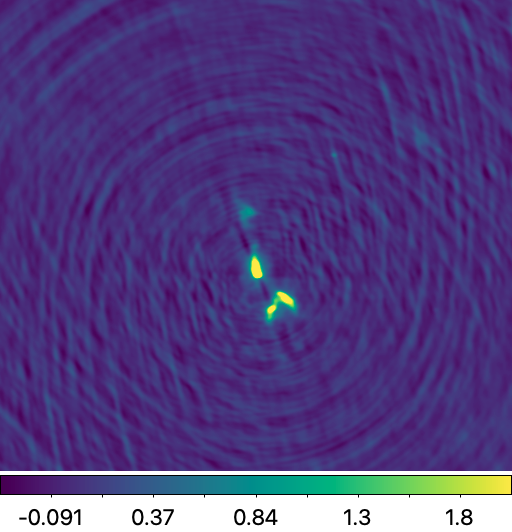} &
 \includegraphics[width=0.23\linewidth]{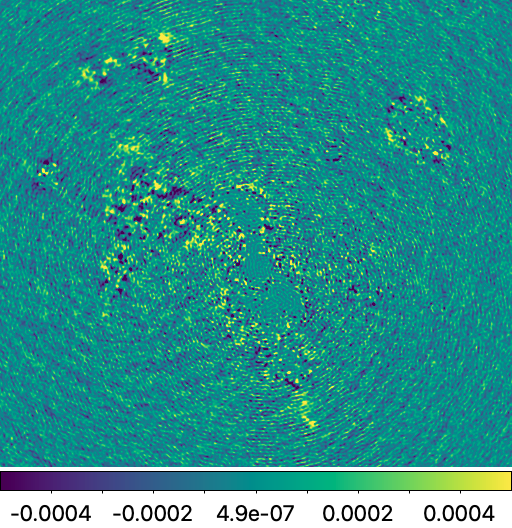} &
 \includegraphics[width=0.23\linewidth]{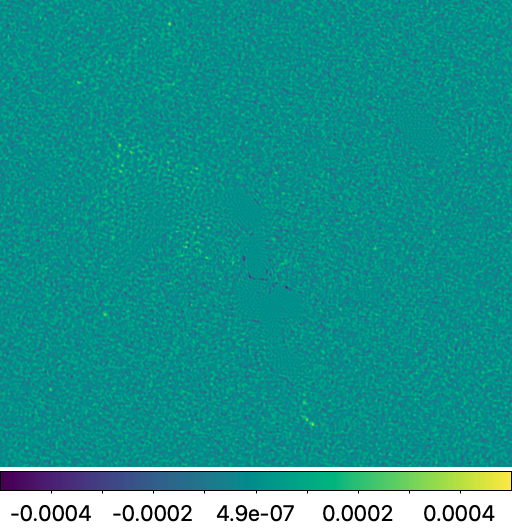} &
 \includegraphics[width=0.23\linewidth]{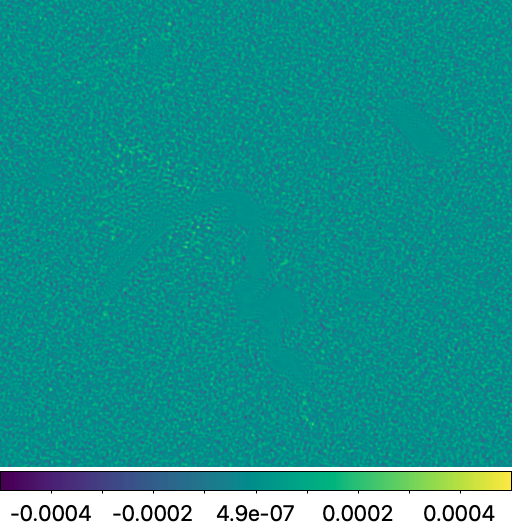}\\
 Dirty $\xb_{\textrm{d}}$& CLEAN: $2.3\times10^{-4}$ & uSARA: $6.{5}\times10^{-5}$ & AIRI: $6.1\times10^{-5}$\\

 \includegraphics[width=0.23\linewidth]{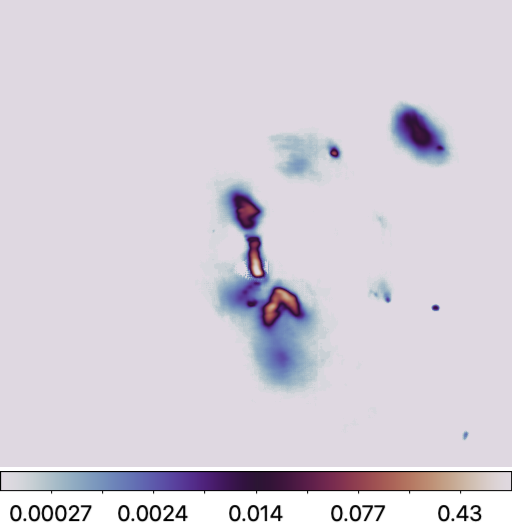} &
 \includegraphics[width=0.23\linewidth]{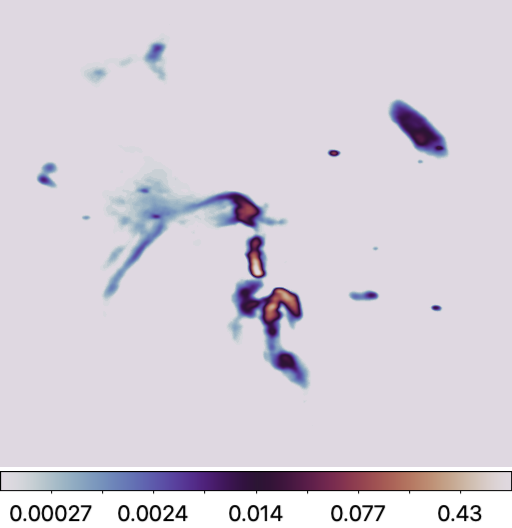} &
 \includegraphics[width=0.23\linewidth]{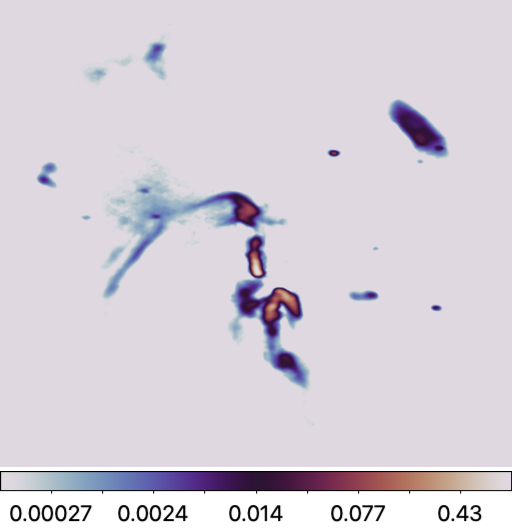} &
 \includegraphics[width=0.23\linewidth]{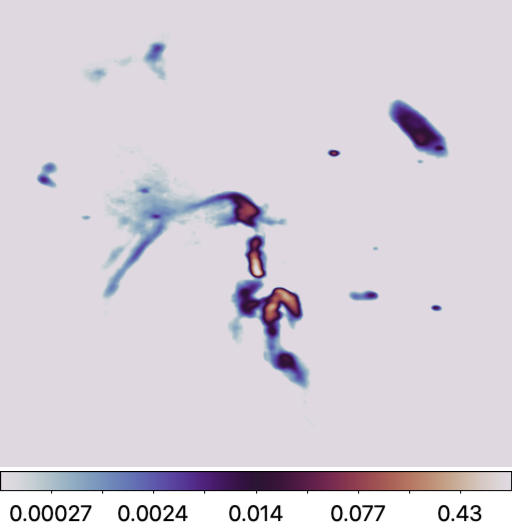}\\
 U-Net: (22.4, 2.3)~dB & R2D2-Net: (33.4, 18.4)~dB & R2D2: (34.0, 21.2)~dB & R3D3: (33.5, 20.8)~dB \\

 \includegraphics[width=0.23\linewidth]{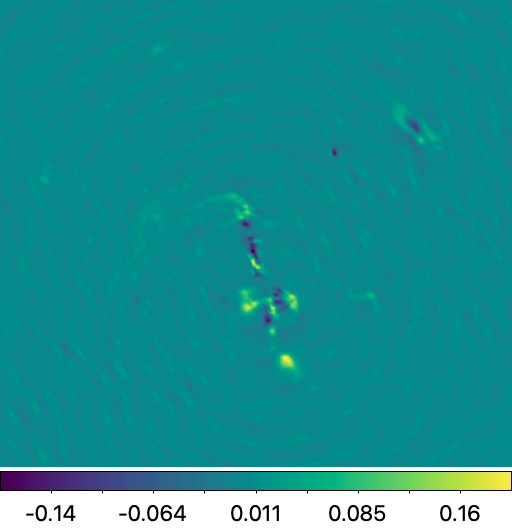} &
 \includegraphics[width=0.23\linewidth]{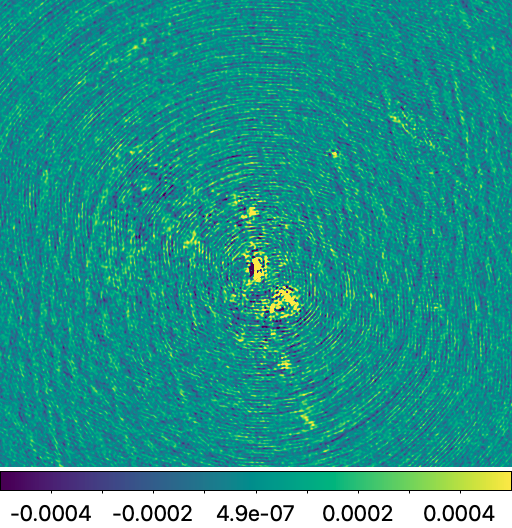} &
 \includegraphics[width=0.23\linewidth]{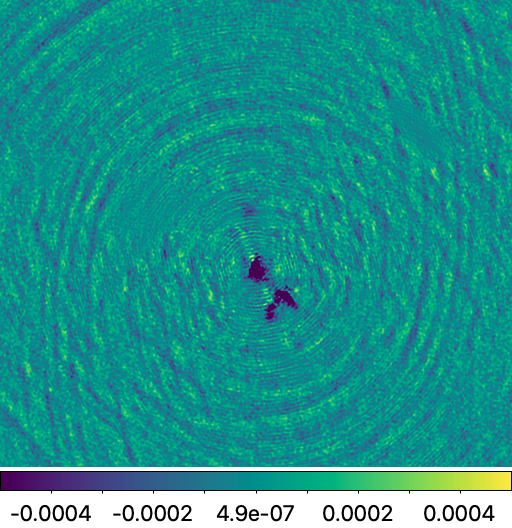} &
 \includegraphics[width=0.23\linewidth]{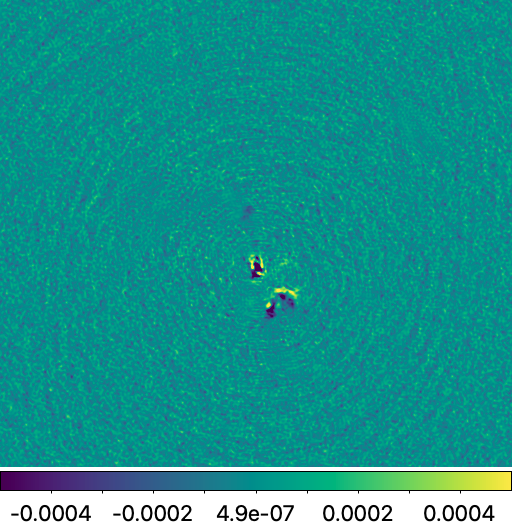}\\
 U-Net: $4.5\times10^{-2}$ & R2D2-Net: $4.6\times10^{-4}$ & R2D2: $6.9\times10^{-4}$ & R3D3: $1.9\times10^{-4}$\\
 \end{tabular}
 \caption{Experiment~I: reconstructions results of the galaxy cluster Abell~2034. The first and third rows show the ground truth image and the estimated images obtained by the imaging algorithms (logarithmic scale). The second and fourth row show the dirty image and the corresponding residual dirty images (linear scale). For R3D3, we showcase the results of its realization with $J=6$ layers in its R2D2-Net components. Values of (SNR, logSNR) are reported below estimated images. ${\overline{\sigma}}_{\textrm{res.}}$ values are indicated below the residual dirty images.} 
 \label{fig:testset1}
\end{figure*}

\begin{figure*}
 \centering
 \setlength\tabcolsep{1pt} 
 \begin{tabular}{cccc}
 \includegraphics[width=0.23\linewidth]{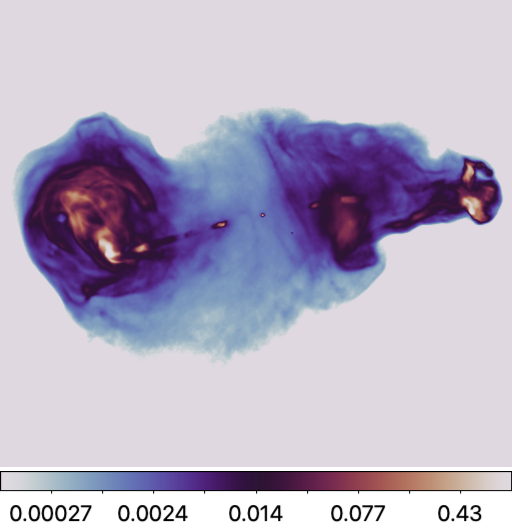} &
 \includegraphics[width=0.23\linewidth]{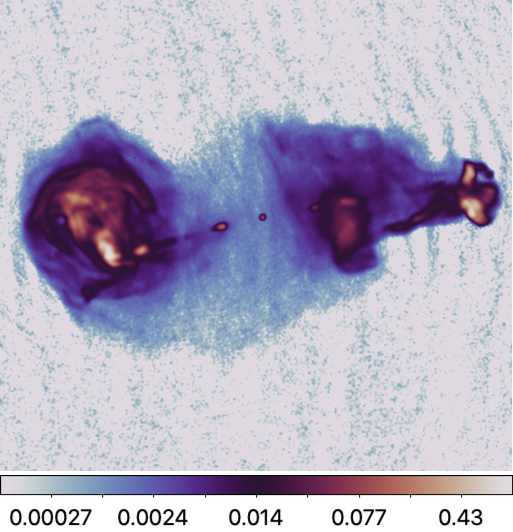} &
 \includegraphics[width=0.23\linewidth]{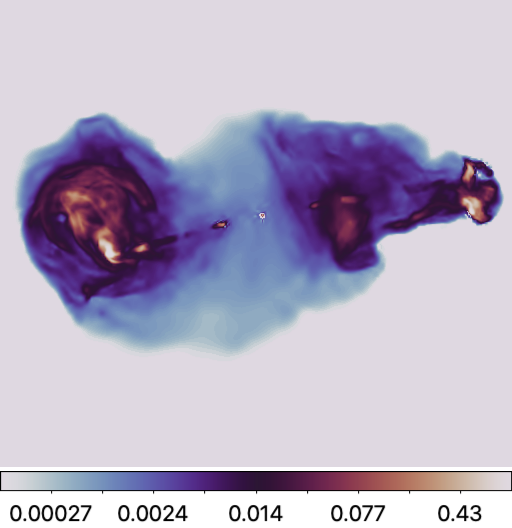} &
 \includegraphics[width=0.23\linewidth]{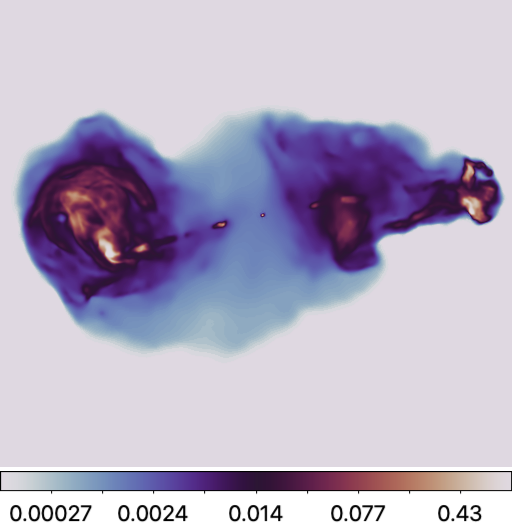}\\
 Ground truth $\xb^{\star}$& CLEAN: (16.0, 15.2)~dB & uSARA: (31.7, 27.4)~dB & AIRI: (35.9, 29.4)~dB \\
 
 \includegraphics[width=0.23\linewidth]{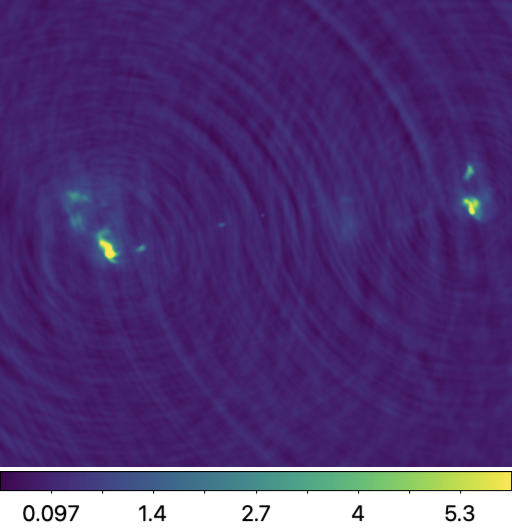} &
 \includegraphics[width=0.23\linewidth]{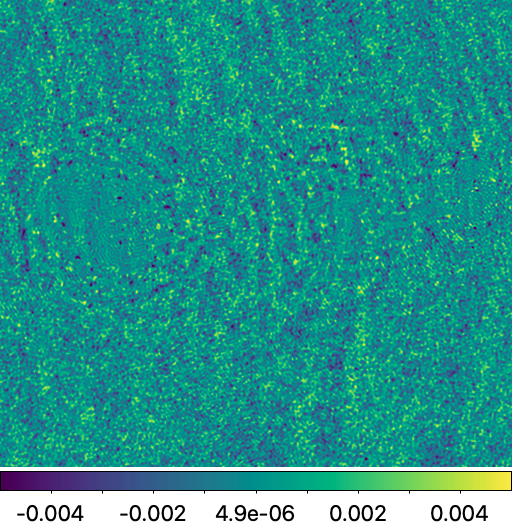} &
 \includegraphics[width=0.23\linewidth]{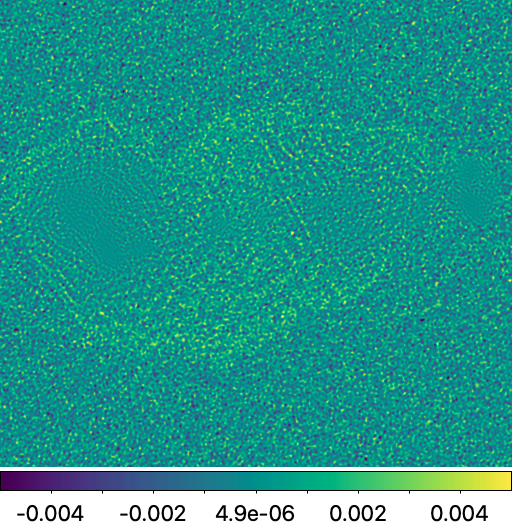} &
 \includegraphics[width=0.23\linewidth]{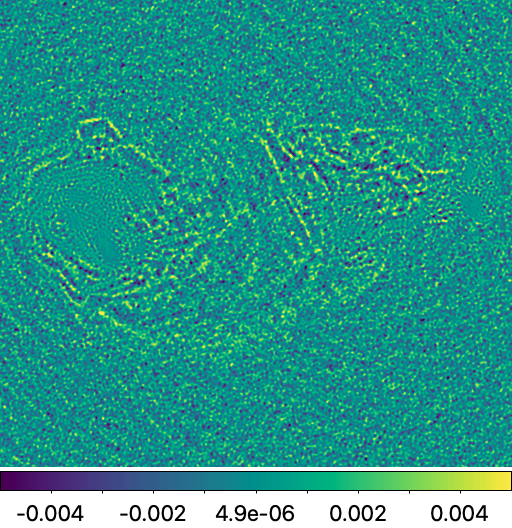}\\
 Dirty $\xb_{\textrm{d}}$& CLEAN: $10.1\times10^{-4}$ & uSARA: $9.1\times10^{-4}$ & AIRI: $10.1\times10^{-4}$ \\

 \includegraphics[width=0.23\linewidth]{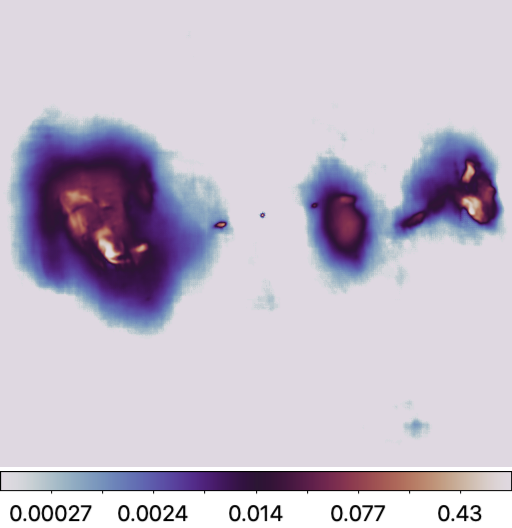} &
 \includegraphics[width=0.23\linewidth]{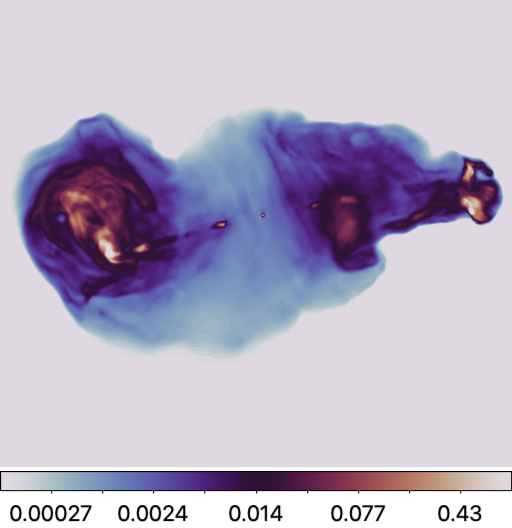} &
 \includegraphics[width=0.23\linewidth]{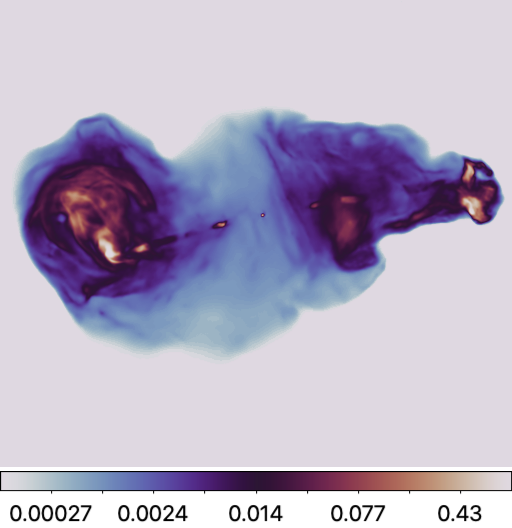} &
 \includegraphics[width=0.23\linewidth]{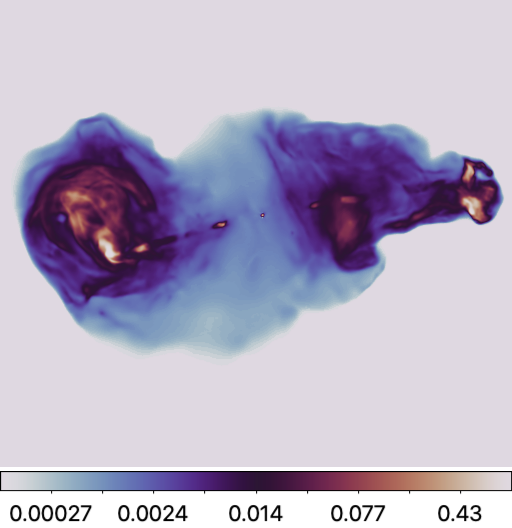}\\
 U-Net: (19.5, 6.2)~dB & R2D2-Net: (37.4, 30.6)~dB & R2D2: (34.7, 32.1)~dB & R3D3: (37.6, 32.3)~dB \\

 \includegraphics[width=0.23\linewidth]{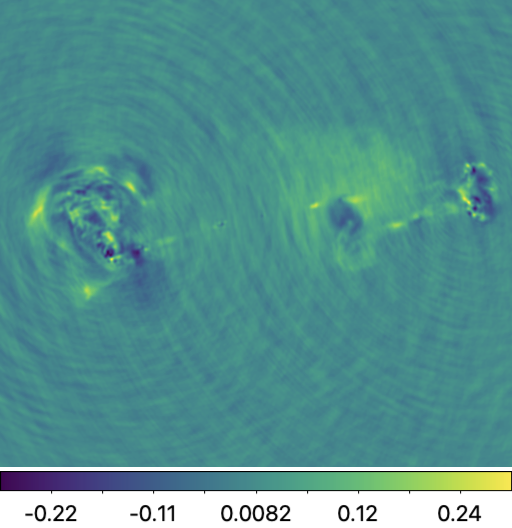} &
 \includegraphics[width=0.23\linewidth]{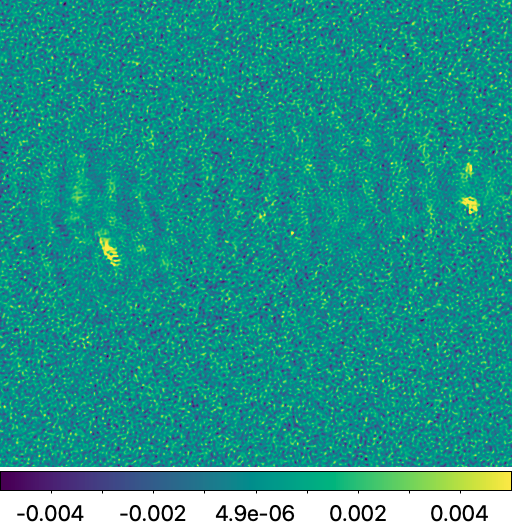} &
 \includegraphics[width=0.23\linewidth]{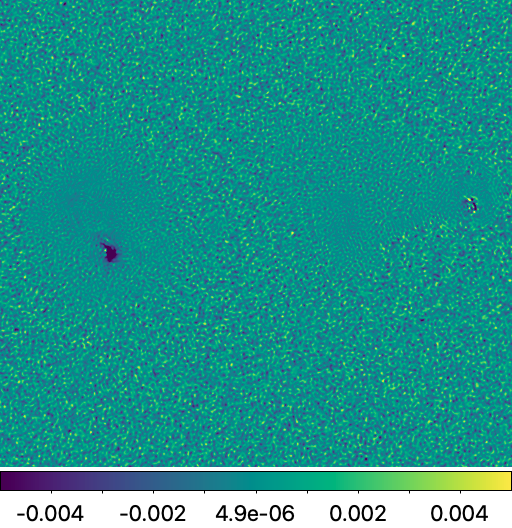} & 
 \includegraphics[width=0.23\linewidth]{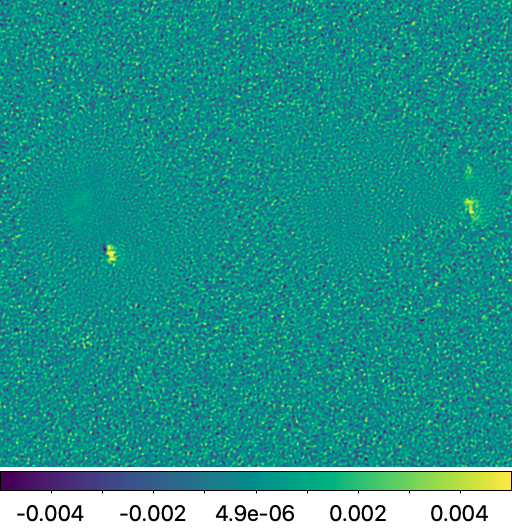}\\
 U-Net: $83.2\times10^{-3}$ & R2D2-Net: $11.0\times10^{-4}$ & R2D2: $10.9\times10^{-4}$ & R3D3: $6.7\times10^{-4}$\\ 
 \end{tabular}
 \caption{Experiment~II: reconstructions results of the radio galaxy 3C353. The first and third rows show the ground truth image and the estimated images obtained by the imaging algorithms (logarithmic scale). The second and fourth row shows the dirty image and the corresponding residual dirty images (linear scale). For R3D3, we showcase the results of its realization with $J=6$ layers in its R2D2-Net components. Values of (SNR, logSNR) are reported below estimated images. The ${\overline{\sigma}}_{\textrm{res.}}$ values are indicated below the residual dirty images.}
 \label{fig:testset2}
\end{figure*}

\begin{figure*}
 \centering
 \setlength\tabcolsep{1pt} 
 \begin{tabular}{cccc}
 \includegraphics[width=0.23\linewidth]{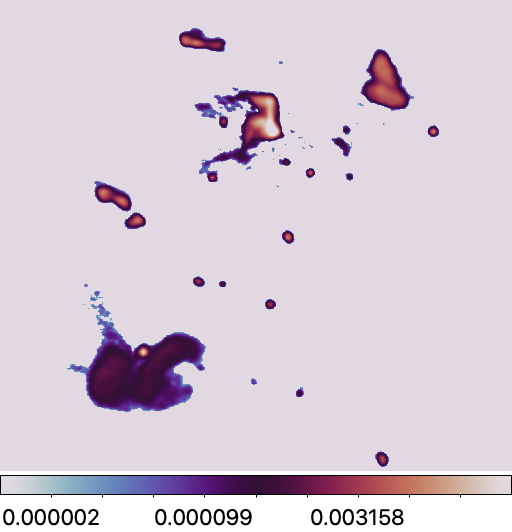} &
 \includegraphics[width=0.23\linewidth]{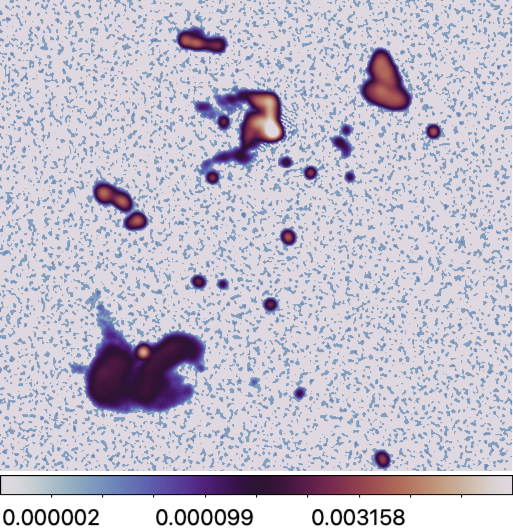} &
 \includegraphics[width=0.23\linewidth]{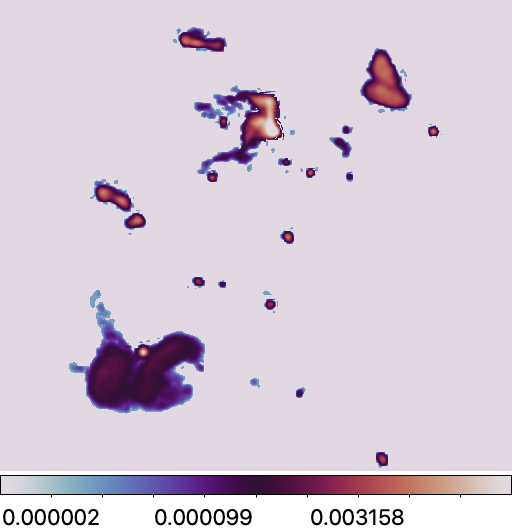} &
 \includegraphics[width=0.23\linewidth]{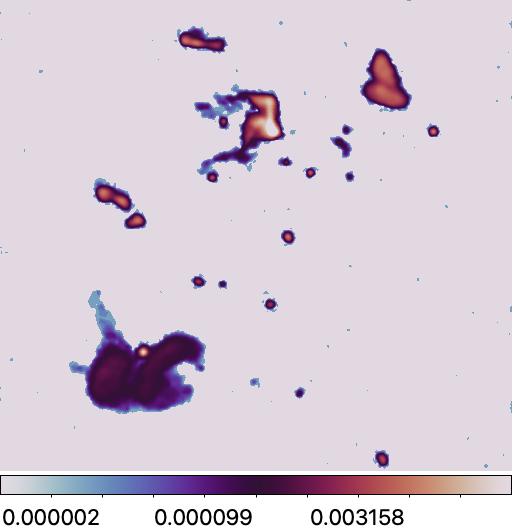}\\
 Ground truth $\xb^{\star}$ & CLEAN: (11.8, 9.5)~dB & uSARA: (30.{3}, 21.1)~dB & AIRI: (30.{2}, 21.1)~dB \\

 \includegraphics[width=0.23\linewidth]{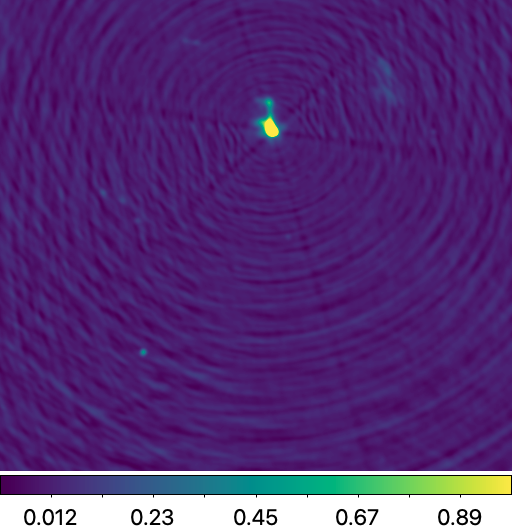} &
 \includegraphics[width=0.23\linewidth]{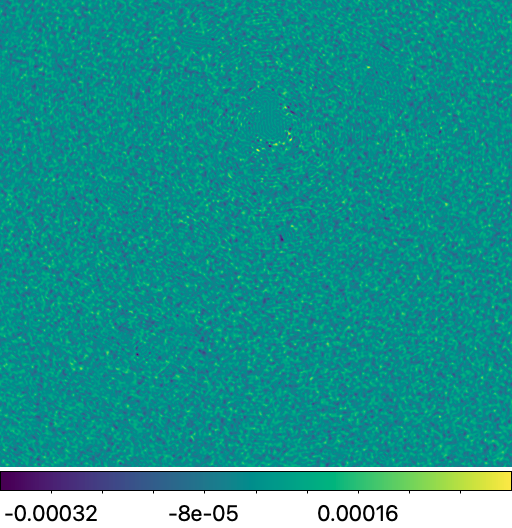} &
 \includegraphics[width=0.23\linewidth]{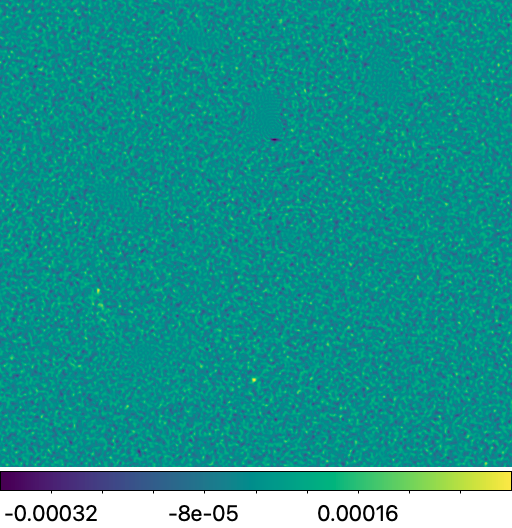} &
 \includegraphics[width=0.23\linewidth]{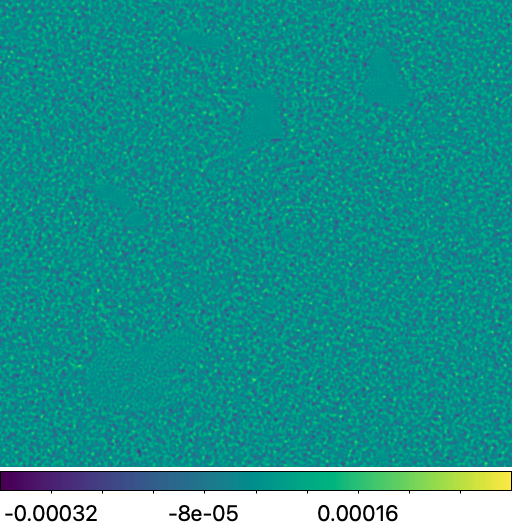}\\
 Dirty $\xb_{\textrm{d}}$& CLEAN: $4.2\times10^{-5}$ & uSARA: $6.{9}\times10^{-5}$ & AIRI: $6.6\times10^{-5}$ \\

 \includegraphics[width=0.23\linewidth]{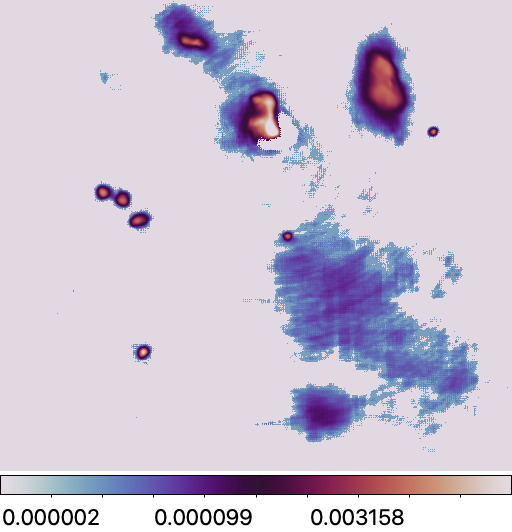} &
 \includegraphics[width=0.23\linewidth]{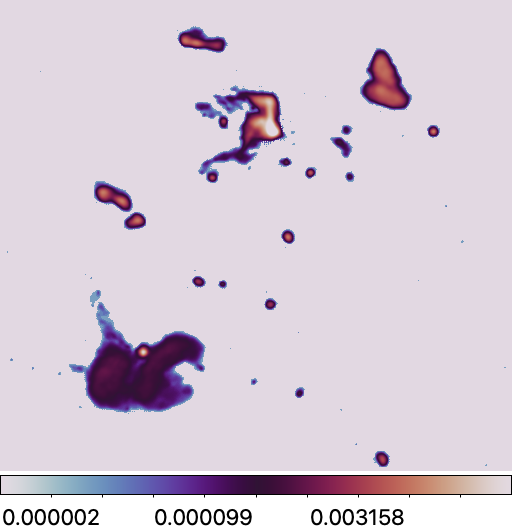} &
 \includegraphics[width=0.23\linewidth]{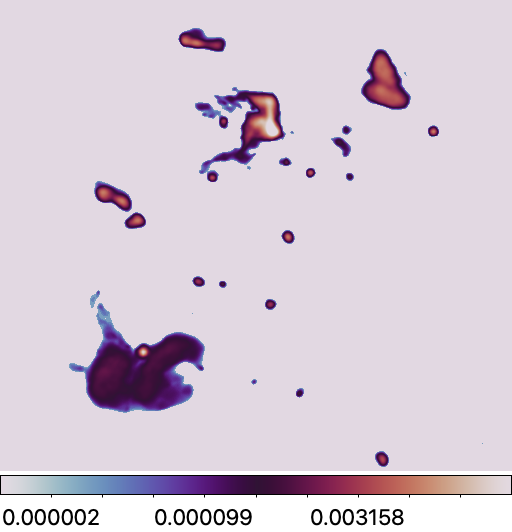} &
 \includegraphics[width=0.23\linewidth]{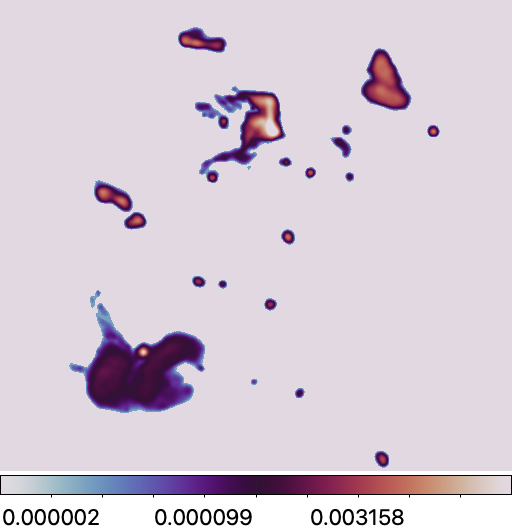}\\
 U-Net: (25.0, 2.0)~dB & R2D2-Net: (31.8, 22.5)~dB & R2D2: (33.4, 24.3)~dB & R3D3: (32.6, 22.9)~dB \\

 \includegraphics[width=0.23\linewidth]{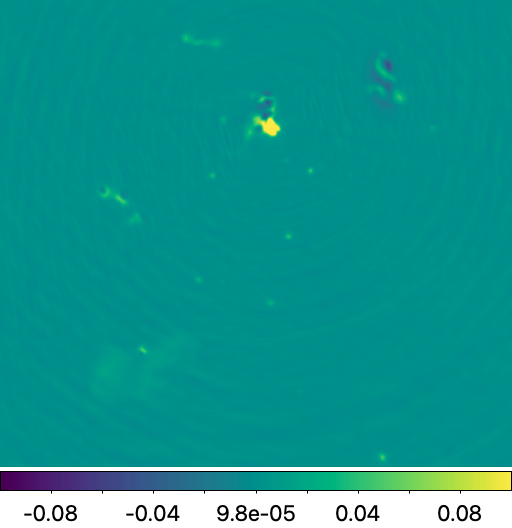} &
 \includegraphics[width=0.23\linewidth]{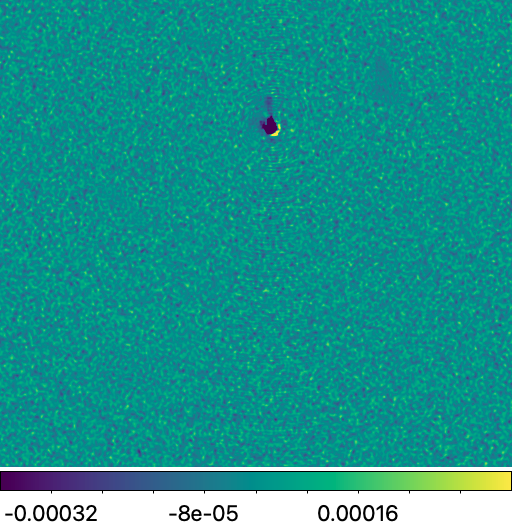} &
 \includegraphics[width=0.23\linewidth]{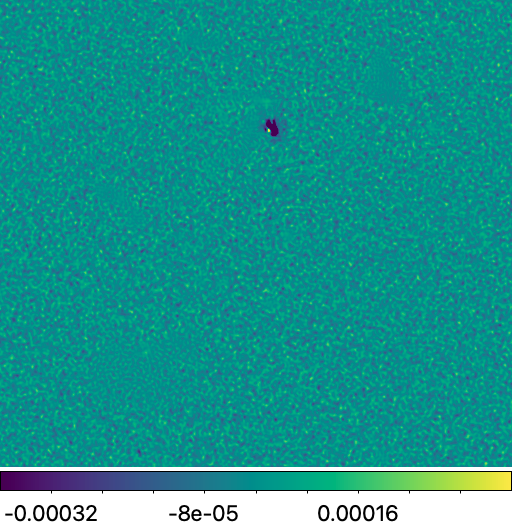} &
 \includegraphics[width=0.23\linewidth]{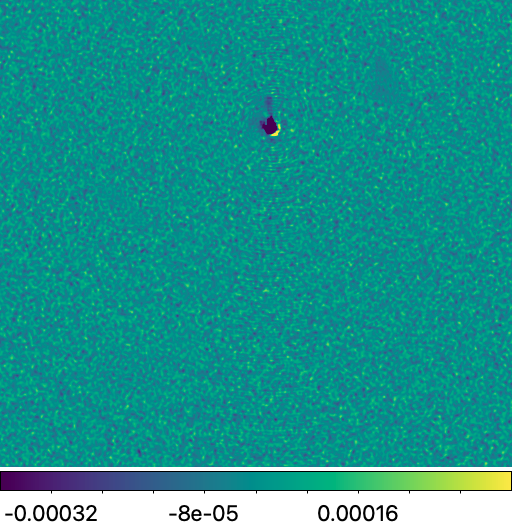}\\
 U-Net: $4.2\times10^{-2}$ & R2D2-Net: $5.9\times10^{-4}$ & R2D2: $2.0\times10^{-4}$ & R3D3: $3.5\times10^{-4}$\\

 \end{tabular}
 \caption{Experiment~III: reconstructions results of the galaxy cluster PSZ2~G165.68+44.01. The first and third rows show the ground truth image and the estimated images obtained by the imaging algorithms (logarithmic scale). The second and fourth row shows the dirty image and the corresponding residual dirty images (linear scale). For R3D3, we showcase the results of its realization with $J=6$ layers in its R2D2-Net components. Values of (SNR, logSNR) are reported below estimated images. The ${\overline{\sigma}}_{\textrm{res.}}$ values are indicated below the residual dirty images.} 
 
 \label{fig:testset3}
\end{figure*}

\begin{figure*}
 \centering
 \setlength\tabcolsep{1pt} 
 \begin{tabular}{cccc}
 \includegraphics[width=0.23\linewidth]{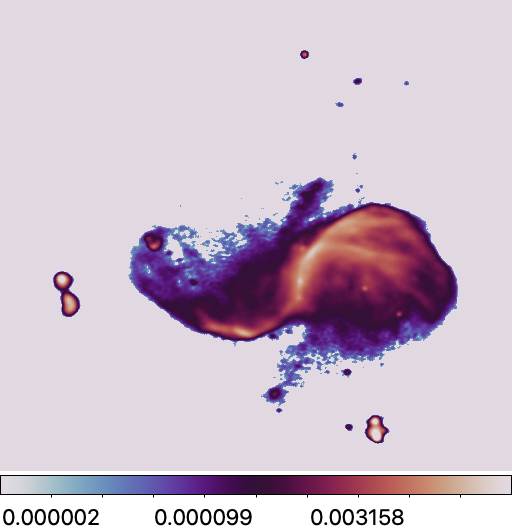} &
 \includegraphics[width=0.23\linewidth]{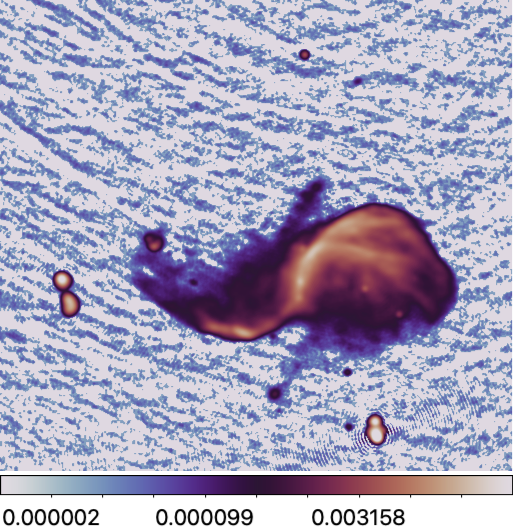} &
 \includegraphics[width=0.23\linewidth]{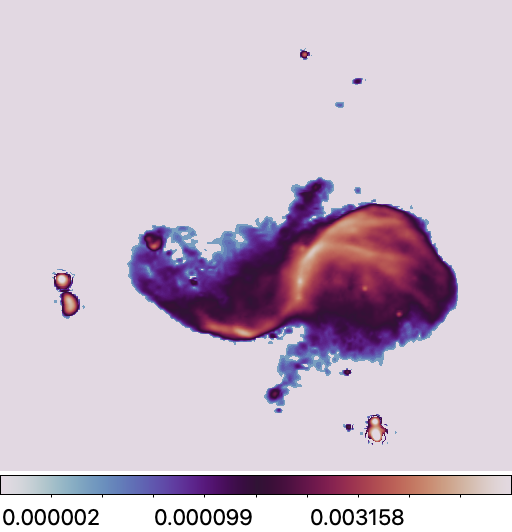} &
 \includegraphics[width=0.23\linewidth]{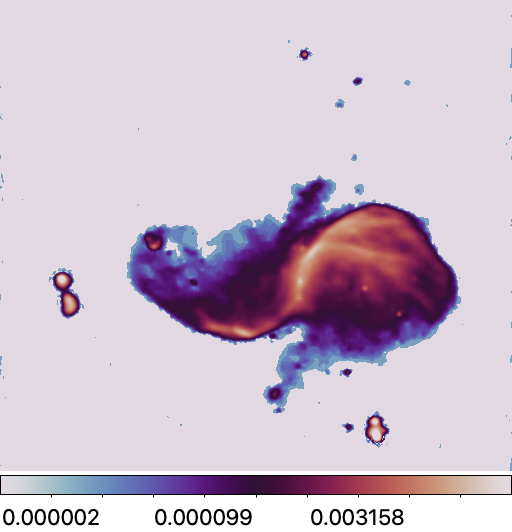}\\
 Ground truth $\xb^{\star}$ & CLEAN: (11.0, 8.8)~dB & uSARA: (30.0, 23.5)~dB & AIRI: (30.8, 23.6)~dB \\

 \includegraphics[width=0.23\linewidth]{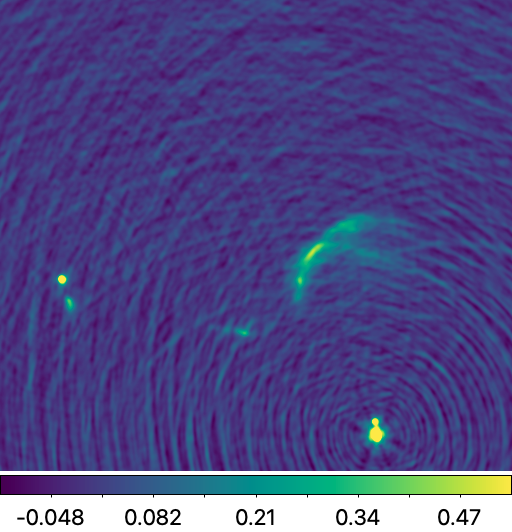} &
 \includegraphics[width=0.23\linewidth]{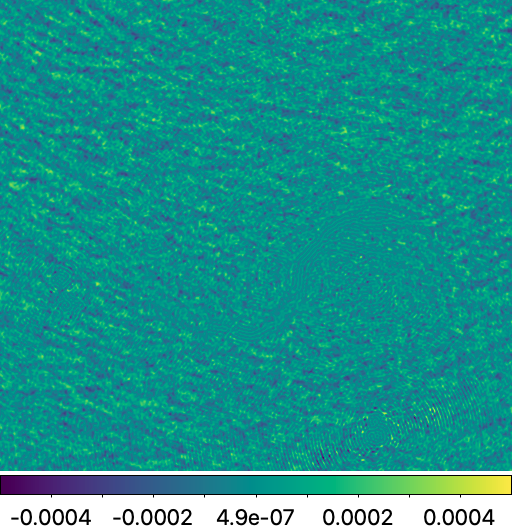} &
 \includegraphics[width=0.23\linewidth]{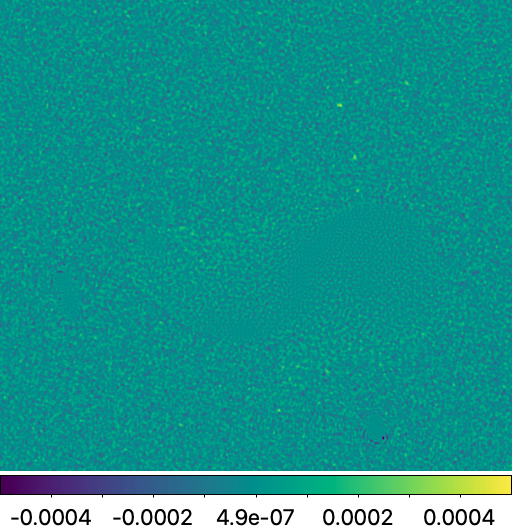} &
 \includegraphics[width=0.23\linewidth]{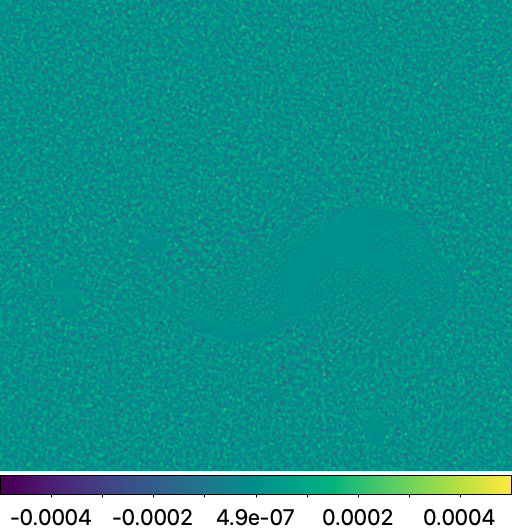}\\
 Dirty $\xb_{\textrm{d}}$& CLEAN: $2.3\times10^{-4}$ & uSARA: $1.1\times10^{-4}$ & AIRI: $1.1\times10^{-4}$ \\

 \includegraphics[width=0.23\linewidth]{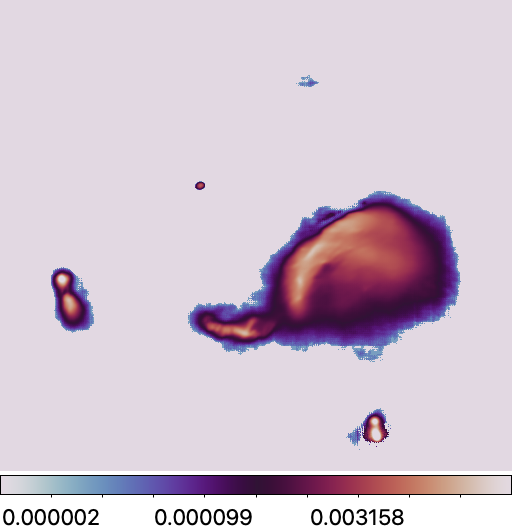} &
 \includegraphics[width=0.23\linewidth]{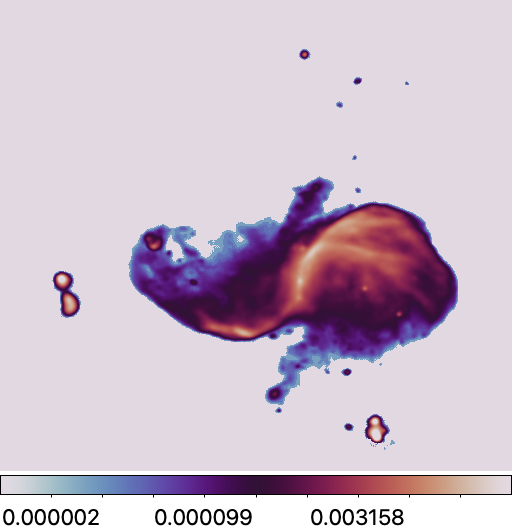} &
 \includegraphics[width=0.23\linewidth]{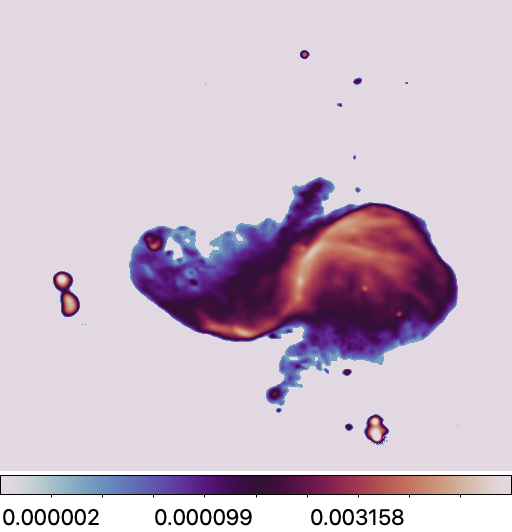} &
 \includegraphics[width=0.23\linewidth]{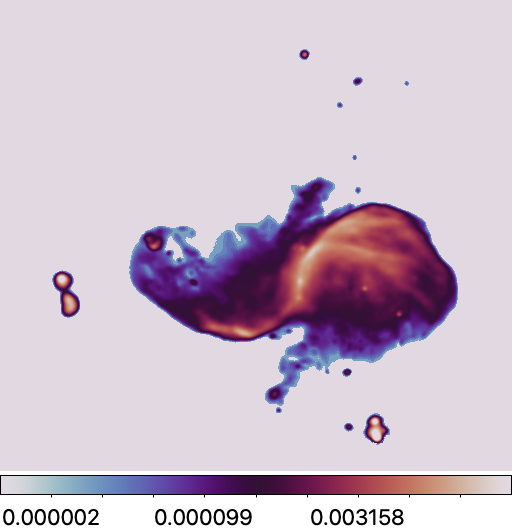}\\
 U-Net: (23.1, 7.6)~dB & R2D2-Net: (34.7, 26.2)~dB & R2D2: (35.8, 26.7)~dB & R3D3: (35,5, 26.7)~dB \\

 \includegraphics[width=0.23\linewidth]{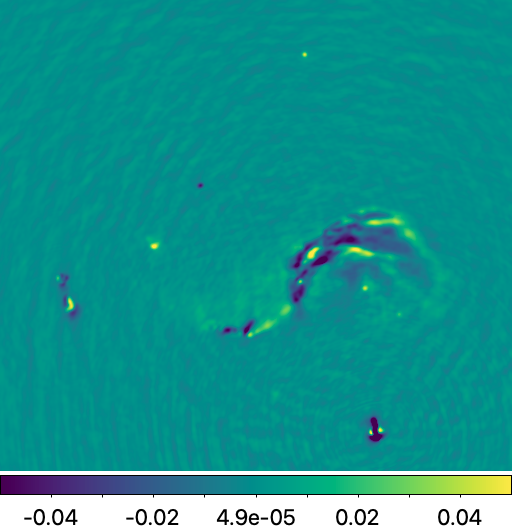} &
 \includegraphics[width=0.23\linewidth]{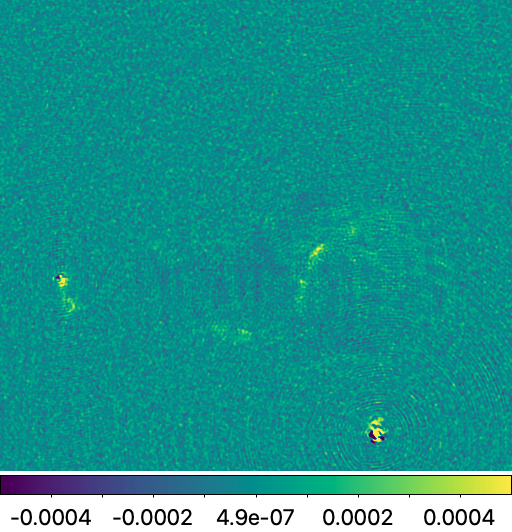} &
 \includegraphics[width=0.23\linewidth]{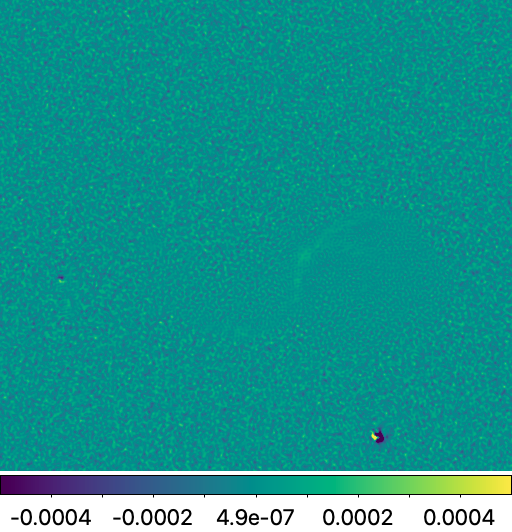} & 
 \includegraphics[width=0.23\linewidth]{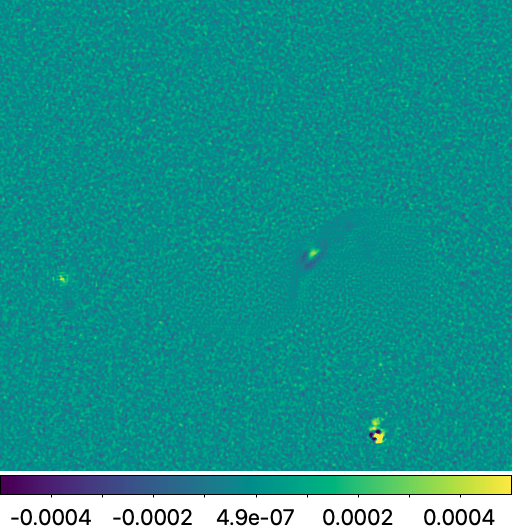}\\
 U-Net: $4.0\times10^{-2}$ & R2D2-Net: ${3.0}\times10^{-4}$ & R2D2: $2.2\times10^{-4}$ & R3D3: ${2.2}\times10^{-4}$\\

 \end{tabular}
 \caption{Experiment~IV: reconstructions results of the radio galaxy Messier~106. The first and third rows show the ground truth image and the estimated images obtained by the imaging algorithms (logarithmic scale). The second and fourth row shows the dirty image and the corresponding residual dirty images (linear scale). For R3D3, we showcase the results of its realization with $J=6$ layers in its R2D2-Net components. Values of (SNR, logSNR) are reported below estimated images. The ${\overline{\sigma}}_{\textrm{res.}}$ values are indicated below the residual dirty images.} 
 \label{fig:testset4}
\end{figure*}

\subsection{Computational performance}\label{subsec:Computational-performance}
Computational details of the different algorithms in terms of the number of iterations and the imaging computational time are presented in Table~\ref{table:results}. A summary of the allocated resources and the programming languages underlying their implementations are also provided. Both incarnations of the proposed algorithm call for a small number of iterations. In contrast, uSARA and AIRI required orders of magnitude more iterations, as expected. As such, the MATLAB implementation of R2D2 and R3D3 delivers reconstructions in seconds, when AIRI and uSARA take around an hour or more. In comparison with CLEAN, both R2D2 and R3D3 call for similar number of passes through the data. Yet, multiscale CLEAN takes over a minute on average, due to its iterative minor cycles, and the slow NUFFT implementation for the relatively small data and image sizes considered in this work. In its hybrid resource allocation setting, the Python version of R2D2 and R3D3 delivers comparable computational time to their MATLAB version. Remarkably, when fully executed on a GPU, it enables faster reconstructions, with an average reconstruction time below 1 second.

Acknowledging the differences in the programming languages and computing hardware of the different algorithms, the reported numbers are indicative only. It is worth noting that both uSARA and AIRI could benefit from additional computational resources for the parallelization of their denoising operators, hence a faster reconstruction. Furthermore, the widely-used WSClean is optimized for large image and data sizes. Also, in the high-dimensional regime of the modern telescopes, uSARA and AIRI demonstrated a substantially reduced computational gap with CLEAN \citep{dabbech2022,wilber23a,wilber23b}. Last but not least, an in-depth investigation of the practical scalability of R2D2 and R3D3 is warranted, particularly given the GPU memory limitations, possibly requiring advanced data and image decomposition approaches at large scale. However, the conclusion stands that the imaging time of R2D2 and R3D3 is significantly lower than AIRI and uSARA thanks to their inherent small number of iterations, and is possibly faster than CLEAN thanks to their DNN inference that is faster than CLEAN's iterative minor cycles.

\section{Conclusions} \label{sec:Conclusions}
We have presented the R2D2 algorithm, a novel deep-learning technique for high-dynamic range imaging in radio astronomy. The R2D2 algorithm is underpinned by a series of end-to-end DNNs trained sequentially, each taking as input the image estimate from the previous iteration alongside its corresponding residual dirty image. The R2D2 reconstruction is formed as the sum of iteratively learned residual images which are outputs of its underpinning DNN series. 

We have provided an in-depth description of the R2D2 paradigm, featuring its two incarnations distinguished by their distinct network components. The first incarnation uses the well-known U-Net architecture, and is simply referred to as R2D2. The second uses a more advanced architecture dubbed R2D2-Net, obtained by unrolling the R2D2 iterative scheme itself. Given its nesting structure, we refer to this incarnation as R3D3. Taking a telescope-specific training approach, R2D2 and R3D3 have been implemented for the VLA. Their imaging precision capability in simulation, across a variety of image and data settings, has been demonstrated against uSARA and AIRI. Owing to their small number of iterations, and fast DNN inference, R2D2 and R3D3 have demonstrated a significant acceleration over uSARA and AIRI. Compared to CLEAN, they require similar number of passes through the data, suggesting comparable computational time independently of their implementation. Propelled by its advanced data model-informed networks, R3D3 has shown similar or higher imaging precision than R2D2 with fewer iterations, potentially enabling further scalability to large data sizes. Of note, the R2D2/R3D3 structure is seen here as parameter-free with the number of required networks set at the training stage, offering an automation advantage over algorithms requiring parameter fine-tuning. Thus, the R2D2 paradigm is opening the door to fast, possibly real-time, high-resolution high-dynamic range RI imaging. 

Future developments are warranted to improve both imaging precision and scalability of the R2D2 algorithm. Firstly, novel core DNN architectures need to be explored to mitigate the residual structure observed in the back-projected residual data. Secondly, a generalized training approach, accommodating yet a wider variety of imaging settings (e.g., varying pixel resolutions and data-weighting schemes, an all-telescopes-encompassing setting) should be investigated. {Thirdly, image-faceting procedures \citep[e.g.,][]{thouvenin2023a} should be deployed at both training and inference, to handle the large image dimensions of interest to SKA and its precursor instruments, especially given the memory limitations of currently available GPUs. The iterative nature of the R2D2 algorithm, in particular its accurate computation of the back-projected data residuals, presents an opportunity for the self-correction of any faceting artifacts.} Finally, advanced data decomposition approaches should be investigated at large scale, for an in-depth study of the practical scalability of the R2D2 algorithm. 

\section*{Data Availability}
R2D2 codes are available alongside the AIRI and uSARA codes in the \href{https://basp-group.github.io/BASPLib/}{BASPLib} code library on GitHub. BASPLib is developed and maintained by the Biomedical and Astronomical Signal Processing Laboratory (\href{https://basp.site.hw.ac.uk/}{BASP}). R2D2 is available in both Python and MATLAB implementations. The VLA-trained R2D2 and R3D3 DNN series are available in both PyTorch and ONNX formats in the dataset \citet{r2d2dnns}.

Images used to generate training, validation, and testing datasets are sourced as follows. Optical astronomy images are gathered from NOIRLab/NSF/AURA/H.Schweiker/WIYN/T.A.Rector (University of Alaska Anchorage). Medical images are obtained from the NYU fastMRI Initiative database \citep{zbontar2018fastmri,knoll2020fastmri}. Radio astronomy images are obtained from the NRAO Archives, LOFAR HBA Virgo cluster survey~\citep{edler2023}, and LoTSS-DR2 survey~\citep{shimwell2022}.

\vspace{4mm}

\software{WSClean \citep{offringa2014,offringa2017},
 Meqtrees \citep{Noordam2010}, PyTorch \citep{paszke2019pytorch}, TorchKbNufft \citep{muckley20};
 }

\begin{acknowledgments} 
The authors thank Yiwei Chen for the insightful discussions on DNN architectures, and Chao Tang for his assistance in building training datasets. The research of AA, AD and YW was supported by the UK Research and Innovation under the EPSRC grant EP/T028270/1 and the STFC grant ST/W000970/1. The research was conducted using Cirrus, a UK National Tier-2 HPC Service at EPCC funded by the University of Edinburgh and EPSRC (EP/P020267/1). The authors thank Adrian Jackson for related support.
\end{acknowledgments}

\bibliography{R2D2}{}
\bibliographystyle{aasjournal}

%
 



\end{document}